\documentclass[prb,aps,floatfix,twocolumn]{revtex4-1}
\usepackage{graphicx}
\usepackage{amsmath, amsthm, amssymb}
\usepackage{color}
\usepackage{pmat}

\usepackage{chngcntr}
\counterwithin*{paragraph}{subsection}

\begin{document}

\title{On Transmission Node Structure in Interacting Systems}

\author{J. D. Barr}

\email{barr@physics.arizona.edu}

\author{C. A. Stafford}
\affiliation{Department of Physics, University of Arizona, 1118 East Fourth Street, Tucson, AZ 85721}

\begin{abstract}

We provide a series of generic results regarding the structure of nodes in the retarded Green's function $G$
of an interacting system, as exemplified by the extended Hubbard model. In particular: (1) due to an incompatibility
between interactions of nearly any form and a precise definition of series propagation, degenerate geometric nodes
are split or lifted by interactions; (2) degenerate nodes generically exist
at the boundary between regimes of node splitting and node lifting and, in the presence of interactions, they require
fine-tuning;
(3) degenerate nodes are highly sensitive to
perturbation and their sensitivity increases with their degeneracy. Moreover, for high degeneracies
there is a tendency toward lifting rather than splitting.

We also propose a characterization of the node structure
in extended Hubbard models at arbitrary filling in terms of either the eigenvalues of $G$, or equivalently, the
roots of a polynomial. This shows that ``Mott nodes'' previously predicted to occur in the transmission spectra
of molecular radicals\cite{justinMottNodePaper} are fundamentally associated with nodes in the eigenvalues of the retarded Green's function
that occur in open-shelled systems. This is
accompanied by a low-energy two-pole approximation wherein each of the eigenvalues of $G$ are mapped onto a Fermi-liquid-like renormalization of the
Anderson model, for which the exact self-energy is provided.

\end{abstract}

\maketitle

\section{introduction}

Absent interactions, the path integral formalism provides an intuitive relationship between the
geometry of a quantum system and the interference effects
exhibited by it. However, the nature of this correspondence in interacting systems is not obvious.

In this article we consider in particular nodes from destructive interference that arise in the retarded Green's function and
consequently the transmission function associated with extended Hubbard models. Such models\cite{hubbardPaper,Ohno64,Castleton02,barrPiEFTPaper} have been used to, for example,
describe transport\cite{dattaMasterEquation,rinconManybodyTransportAnnulenes,barrPiEFTPaper,justinManybodyPaper,justinSupernodePaper,2007bohr,meirWingreenPaper,meirWingreenPaper2,meirWingreenTimeDependentReference} through molecular junctions
and mesoscopic systems. In this context there
is a great deal of experimental and theoretical evidence that most nodes present within noninteracting models
persist in the presence of interactions.\cite{BDTNodePaper1,BDTNodePaper2,BDTNodePaper3,BDTNodePaper4,BDTNodePaper5,
BDTNodePaper8,BDTNodePaper9,BDTNodePaper10,BDTNodePaper11,BDTNodePaper12,BDTNodePaper13,justinManybodyPaper,connectivityConductanceExperimentPaper}

  \begin{figure}
	\centering
	\includegraphics[scale=1.00]{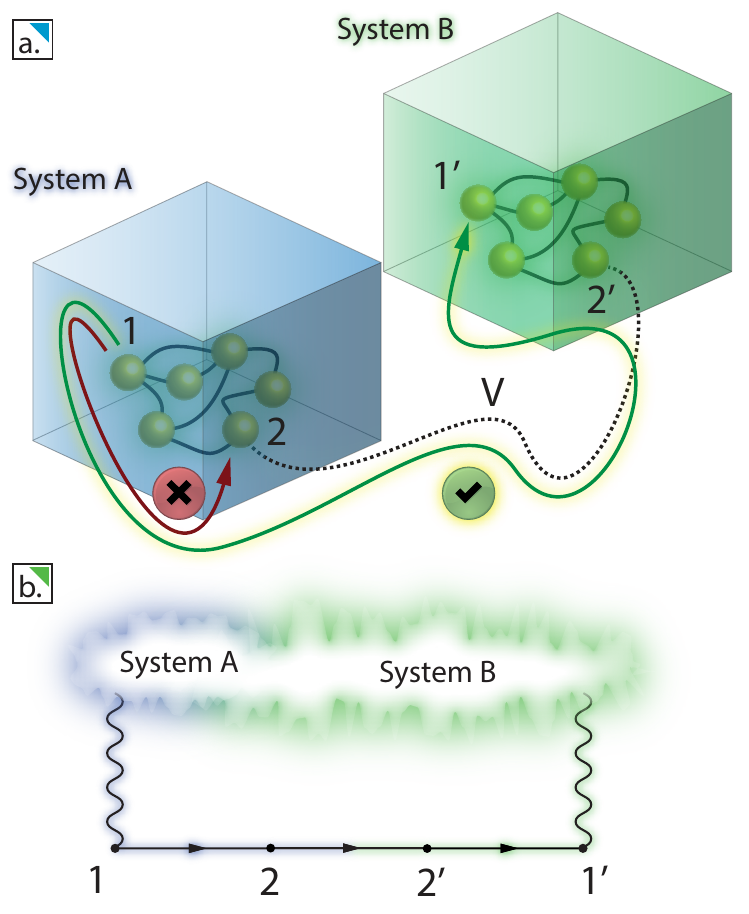}
\caption{(a) Two identical systems, $A$ and $B$, linked by a matrix element $V$ between orbitals $2$ and $2'$. Even with interactions only \emph{within} each
system, a definition of series propagation is fundamentally broken by the processes depicted in (b). This can cause dramatic changes in the low-energy node structure
of the propagator, e.g. permitting transport between $1$ and $1'$ when it is forbidden between $1$ and $2$.} 
\label{series_transport_boxes}
\end{figure}
However, in contrast to this, degenerate nodes present in
interacting systems were previously observed to be split by Coulomb interactions within many-body calculations.\cite{justinNodeSplittingPaper}$^,$\footnote{After
preparation of this manuscript, we became aware via private correspondence that J. Bergfield has recently observed node splitting 
within a Hubbard model of a molecular junction in unpublished work.}
Here we propose that the foregoing observation is the result of an inherent incompatibility
between the presence of interactions and a definition of series propagation that can be formulated
diagrammatically. Remarkably, we find violations of this definition cause large disruptions
of degenerate nodes even when there are no interactions
between the units arranged in series. The reason is fundamental: Amplitudes in interacting systems are not
expressible as a sum of \emph{geometric} Feynman paths. All possible field configuration \emph{histories} contribute
coherently to a propagator, and degenerate nodes are found to be strikingly sensitive to this (Figure \ref{series_transport_boxes}).

We also show for the first time that degenerate nodes do exist in interacting systems. However, in light
of the aforementioned breakdown of series propagation, we find that they generically require fine-tuning and
that their sensitivity to perturbation and tuning increases with their order. Moreover, we provide the first reports
that degenerate nodes can be lifted by interactions instead of split, that they generically lie at the boundary between regimes of splitting
and lifting, that this splitting or lifting is an ill-conditioned function of parameters in the Hamiltonian, and that for high degeneracy
there is a tendency toward lifting.

To provide a simple conceptual framework for understanding the node structure in extended Hubbard models,
we also develop a simple but remarkably accurate functional form
for the eigenvalues of the low-energy retarded Green's function that qualitatively characterizes
the node structure of the models we consider. It is reminiscent of 
two-pole approximations that have been applied to the Hubbard model,\cite{hubbardPaper,rothPaper} and it is found to be equivalent
to mapping each eigenvalue of the retarded Green's function
onto the Green's function of a Fermi-liquid-like\cite{landauFermiLiquidPaper} renormalization of the Anderson model.\cite{andersonAndersonModel}

Using this, we show that so-called ``Mott nodes'', reported previously
in theoretical predictions of molecular transmission spectra\cite{justinMottNodePaper}
and in the Hubbard model,\cite{mottNodePaper} are associated with nodes in the eigenvalues of the retarded Green's function
in extended Hubbard models. We note
that such nodes also appear at high energies, and identify them as being generically due to the
interference of field configuration histories. In this context, ``eigenvalue nodes''
responsible for Mott nodes are interpreted as being due to the destructive interference of particle-like
and hole-like processes mediated by the \emph{same} single-particle orbital. This
refines earlier work\cite{justinMottNodePaper} advancing the interpretation that
the Mott node itself is directly due to particle-hole interference, which, while true in the
context previously considered, turns out to be a special case.

Finally, we also cast the nodes of the retarded Green's function
as the roots of a polynomial, which provides a formal connection
between the perturbation of nodes and the perturbation of the
coefficients of a polynomial. While this formalism gives the nodes of the Green's function
exactly at all energies, we find it practical to develop a reduced-order polynomial
that characterizes low-energy node structure. In this context, split and lifted degenerate
nodes correspond respectively to the real and complex roots of this polynomial.

The organization of this article is as follows: In Section \ref{sec:seriesPropagation} we develop a definition of series propagation
and remark briefly upon the connection between propagation in an isolated system and transmission through it.
In Section \ref{sec:topology}, we provide an overview of the relationship between the breakdown of this definition of series propagation
and the disruption of degenerate nodes in the presence of Coulomb interactions. In Section \ref{sec:electroncStructure},
we describe a connection between electronic structure and node structure in extended Hubbard models that
characterizes low-energy nodes in such systems at arbitrary filling. This is then used to provide insight into
a few generic properties of the node structure in interacting systems.

Many of the results that are described in Sections \ref{sec:topology} and \ref{sec:electroncStructure} are based upon
a detailed series of case studies of nodes in simple extended Hubbard models. For the sake of clarity, these
do not appear in the main text and instead have been organized in Appendix \ref{sec:caseStudies}.
Similarly, the technical details regarding the connection between nodes in the transmission function
and nodes in the retarded Green's function appear in Appendix \ref{sec:transport}. Finally, a derivation of the two-pole approximation
for the eigenvalues of the Green's function used in Section \ref{sec:electroncStructure} appears in Appendix \ref{sec:spectralDecomposition}.

\label{sec:formalism}
\begin{widetext}
\section{A definition of series propagation}
\label{sec:seriesPropagation}
\label{sec:greensFunctionDefinition}

Consider the system depicted in Figure \ref{series_transport_boxes} consisting of two interacting systems $A$ and $B$
linked by a single matrix-element $V$.
For the sake of concreteness, we take an extended Hubbard model\cite{hubbardPaper} with spin and particle-hole symmetry for the interactions, in
which case the Hamiltonian of the composite system $A + B$ is of the form:
\begin{align}
	\label{generalSeriesTransportHamiltonian}
\mathcal{H}^{A + B} = & \underbrace{-\sum_{i,j,\sigma} H^A_{ij} d^\dagger_{i\sigma} d_{j\sigma} + \frac{1}{2}\sum_{i,j} U^A_{ij} \rho_i \rho_j }_{\textrm{System $A$}} \nonumber \\
& \underbrace{-\sum_{i,j,\sigma} H^B_{ij} c^\dagger_{i\sigma} c_{j\sigma} + \frac{1}{2}\sum_{i,j} U^B_{ij} \eta_i \eta_j  }_{\textrm{System $B$}}  \nonumber \\
& \underbrace{-\sum_{\sigma} \left( V c^\dagger_{n\sigma} d_{m\sigma} + \textrm{H.c.}\right) + \frac{1}{2} \sum_{i,j} U^{AB}_{ij} \rho_i \eta_j}_{\textrm{Coupling}}
\end{align}
where $\rho_i = \sum_{\sigma} d^\dagger_{i\sigma} d_{i\sigma} - 1$ and $\eta_i = \sum_{\sigma} c^\dagger_{i\sigma} c_{i\sigma} - 1$.
Here $c^\dagger_{n\sigma}$ and $d^\dagger_{n\sigma}$ respectively create electrons with spin $\sigma$ on the $n$th site of systems $A$ and $B$ .

The amplitude for a particle added to the $\alpha$th orbital of system $A$ at time $0$ to be observed in the $\beta$th
orbital of system $B$ at time $t$ is proportional to a Green's function $G_{\beta\alpha}(t, 0) = -i\hbar \Theta(t) \langle
d_\beta(t) d^\dagger_\alpha(0) \rangle$. We define series propagation in this context via the requirement that
$G_{\beta \alpha}$ be expressible as a coherent sum over Feynman paths,\cite{1948feynman} each corresponding to a process wherein a
particle propagates back and forth through each system in series, eventually ending up in the $\beta$th orbital of
system $B$. The most general expression for $G_{\beta\alpha}$ under these conditions is:
\begin{align}
	\label{series_propagation_definition}
G_{\beta\alpha}(t, 0) = & \underbrace{ \int_0^t d\tau_1 \; G^{A}_{n \alpha}(\tau_1, 0) \Delta G^{B}_{\beta m}(t, \tau_1)}_{\textrm{Direct transmission}} \nonumber \\
+ & \; \underbrace{\int_0^t d\tau_1 \int_{\tau_1}^t d\tau_2 \int_{\tau_2}^t d\tau_3 \; G^{A}_{n \alpha }(\tau_1, 0) \Delta G^{B}_{mm}(\tau_2, \tau_1) \Delta^* G^{A}_{nn}(\tau_3, \tau_2) \Delta G^{B}_{\beta m}(t, \tau_3)}_{\textrm{Transmission mediated by reflections at times $\tau_1$ and $\tau_2$}} \nonumber \\
+ & \; \ldots
\end{align}
where $\Delta$ is an amplitude associated with hopping between $A$ and $B$, and $G^A$ and $G^B$ are amplitudes associated with propagation within $A$ and $B$.
These do not need to be Green's function associated the \emph{isolated} systems, though they can be.
We merely assume all the amplitudes are causal in the sense that $G_{ij}(t, 0) = 0$ for $t < 0$.

In this case, time translation invariance implies:
\begin{align}
	G_{\beta \alpha}(t, 0) = &  \int_{-\infty}^{\infty} d\tau_1 \, G^{A}_{n \alpha}(\tau_1) \Delta G^{B}_{\beta m}(t - \tau_1) \nonumber \\
+ & \; \int_{-\infty}^{\infty} d\tau_1 \, d\tau_2 \, d\tau_3 \, G^A_{n \alpha}(\tau_1) \Delta G^B_{mm}(\tau_2 - \tau_1) \Delta^* G^A_{nn}(\tau_3 - \tau_2) \Delta G^B_{\beta m}(t - \tau_3) \nonumber \\
+ & \; \ldots
\end{align}
which is a Dyson equation\cite{dysonEquationPaper} with the self-energy $\Sigma_{ij}(\tau) = \Delta \delta(\tau)\delta_{in}\delta_{jm}$.
Applying the convolution theorem then yields a geometric series in the energy domain:
\begin{align}
	G_{\beta \alpha} = & \; G^A_{n \alpha } \Delta G^B_{\beta m} + G^A_{n \alpha } \Delta G^B_{mm}\Delta^* G^A_{nn} \Delta G^B_{\beta m} + \; \ldots \nonumber \\
= & \; \frac{G^A_{n\alpha}\Delta G^B_{\beta m}}{1 - G^B_{mm} \Delta^* G^A_{nn}\Delta } = \frac{G^A_{n\alpha} V G^B_{\beta m}}{1 -G^A_{nn} G^B_{mm} |V|^2 } \label{pureSeriesEnergyDomain}
\end{align}
The last equality follows from the condition that the hopping amplitude $\Delta$, which has not been specified till now, is equal to $V$. This ensures that the
foregoing definition gives the exact result for noninteracting systems, as the self-energy $\Sigma$ above is then the tunneling self-energy\cite{meirWingreenPaper2} associated with $V$.
For clarity, the energy dependence of the amplitudes $G_{ij}$ is left as implicit. 

We note here that, although the
intuitive case of $G_{\beta \alpha}(t, 0) = -i\hbar \Theta(t) \langle d_\beta(t) d^\dagger_\alpha(0) \rangle$ was
considered first to formulate our definition of series propagation, the preceding applies equally to the retarded Green's
function from the Keldysh formalism:\cite{keldyshPaper} $G_{\beta \alpha}(t, 0) = -i\hbar \Theta(t) \langle \{ d_\beta(t) ,
d^\dagger_\alpha(0) \} \rangle$. Transport related quantities are elegantly expressed via this quantity, as it combines
the amplitudes for both particle-like and hole-like processes.\cite{jauhoBook, meirWingreenPaper, meirWingreenPaper2}

Thus, throughout the remainder of this work we consider only the retarded Green's function, though the preceding is not specific to it.
The precise relationship between the retarded Green's function of a system and the associated elastic transmission function is given in Appendix \ref{sec:transport}.
We also note that when one of the systems is a set of metallic electrodes and bare Green's functions are taken for $G^{A,B}$,
the definition presented here reduces to the so-called elastic cotunneling approximation that has been studied before.\cite{1991groshev, 1990averin, 2004glazman}

We now consider the relationship between this definition of series propagation and node structure in interacting systems.
\end{widetext}

\section{Topology and node structure: The breakdown of series propagation}
\label{sec:nodeStructureOverview}
\label{sec:topology}

Eq. \eqref{pureSeriesEnergyDomain} implies that a node at energy $E$ in $G^{A}_{n \alpha}$ is sufficient
to cause a node at the same energy in $G_{\beta \alpha}$ provided $G^{B}_{\beta n}$ is bounded in the vicinity of $E$. Equivalently,
within the foregoing formulation of series transmission, if a virtual particle with a given energy can not propagate from $\alpha$ to $n$, it can
not propagate from $\alpha$ to $\beta$ by way of
coupling between $n$ and $m$. In particular, if $G^A_{n \alpha}$ and $G^B_{\beta m}$ both vanish according to a power law, i.e. $G \propto (E - E_0)^\eta$,
then Eq. \eqref{pureSeriesEnergyDomain} implies the amplitude $G_{\alpha \beta}$ exhibits a degenerate node, vanishing as $(E - E_0)^{2\eta}$.

\begin{figure}
	\centering
	\includegraphics[scale=0.98]{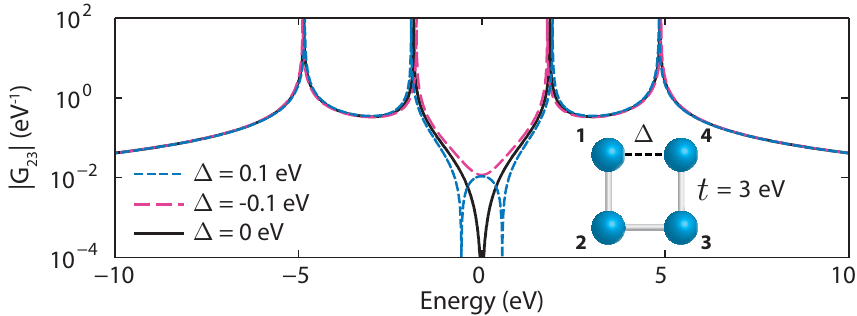}
	\caption{The element $G_{23}$ of the retarded Green's function for the depicted tight-binding model ($t_{12} = t_{23} = t_{34} \equiv t = 3$ eV,
	$t_{14} \equiv \Delta$). This system exhibits a degenerate node at $E = 0$ when $\Delta = 0$ eV (black), but for $\Delta = \pm 0.1$ eV the node is respectively
	lifted or split. Although in this case there are no interactions,
	we find that degenerate nodes generically sit at the boundary between node splitting and lifting.\label{noninteracting_splittingLifting}}
\end{figure}

\begin{figure}
	\centering
	\includegraphics[scale=1.0]{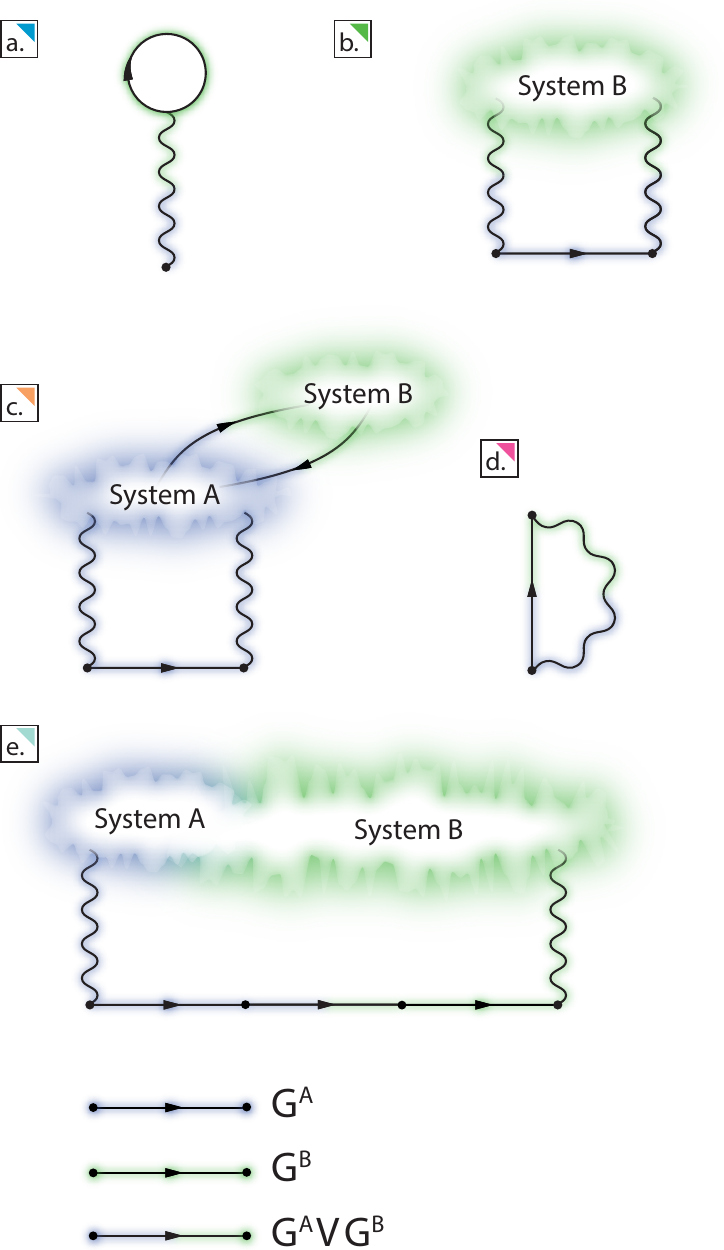}
\caption{Diagrams depicting processes that are inconsistent with a definition of pure series propagation between system A (blue) and system B (green). Diagrams (a) - (c)
are consistent with a weak definition of series propagation, whereas (d) and (e) are not. We find the presence of either (d) or (e) destroys degenerate nodes otherwise expected
on the basis of series propagation.\label{feynman_diagrams_generic}}

\end{figure}

With interactions these properties do not hold because the foregoing definition of series propagation, and consequently Eq. \eqref{series_propagation_definition}, breaks down.
This is true even absent interactions between systems A and B.
In the case studies presented in section \ref{sec:caseStudies}, we find that \emph{even when there are no Coulomb interactions between the systems}, and \emph{even
when $G^A$ and $G^B$ are dressed exactly by the coupling of $A$ to $B$}, violations the foregoing definition of series propagation dramatically
alter the structure of degenerate nodes at, for example, molecular energy scales. Ultimately, the reason is fundamental: Amplitudes in interacting systems are not expressible as a sum of \emph{geometric} Feynman paths.
All possible field configuration \emph{histories} contribute
coherently to a propagator. Thus, the factorization defined by Eq. \eqref{series_propagation_definition} omits processes like those represented by the Feynman diagrams
depicted in Figure \ref{feynman_diagrams_generic}:

Diagrams of class (a) and (b) are present only when there are long-ranged Coulomb interactions between $A$ and $B$, and can be interpreted as dressing the Green's
function in Eq. \eqref{series_propagation_definition}. Diagrams of class (c) can be viewed in the same manner but exist even absent intersystem interactions. The
nonlocal exchange in (d) does require long-ranged Coulomb interactions, but can not be treated by dressing $G^A$. It is formally equivalent to introducing new matrix
elements between $A$ and $B$ that alter their mutual connectivity. The final class (e) is inconsistent with series propagation in the purest sense -- it exists generically,
can not be treated within any formalism that involves only single-particle Green's functions for $A$ and $B$, and does not arise from Coulomb interactions between
the systems.

In the case studies presented in Appendix \ref{sec:caseStudies}, we find that for two identical
systems arranged in series, the presence of diagrams of the form (d) or (e) is necessary
and sufficient for the disruption of degenerate nodes that would otherwise be expected on the basis of series propagation. When present, diagrams
of the form (d) usually correspond to self-energies that are quantitatively larger than those associated with the higher order diagrams (e). However,
we find either (d) or (e) is capable of imparting very large changes to interference effects otherwise expected on the basis of series propagation. Under
some circumstances, we find they compete and lead to novel node structure.

\begin{figure}
	\centering
	\includegraphics[scale=1.0]{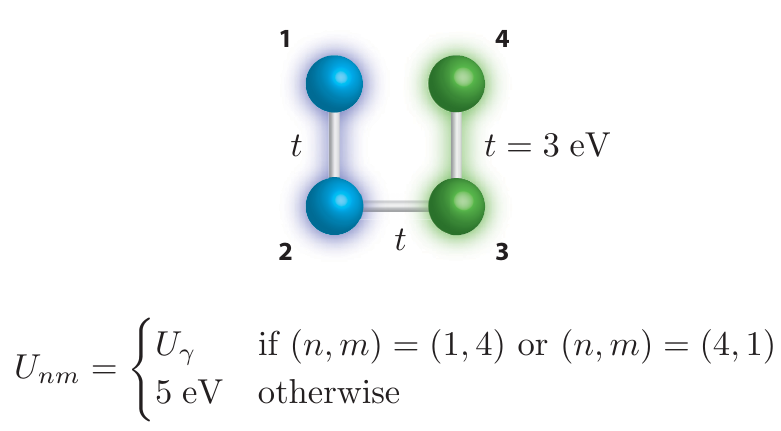}
\caption{The geometry and extended Hubbard parameters associated with the interactiong supernode depicted in Figure \ref{fourSitesTuning}.}
\label{fourSitesTuningGeometry}
\end{figure}
\begin{figure}
	\centering
	\includegraphics[scale=1.0]{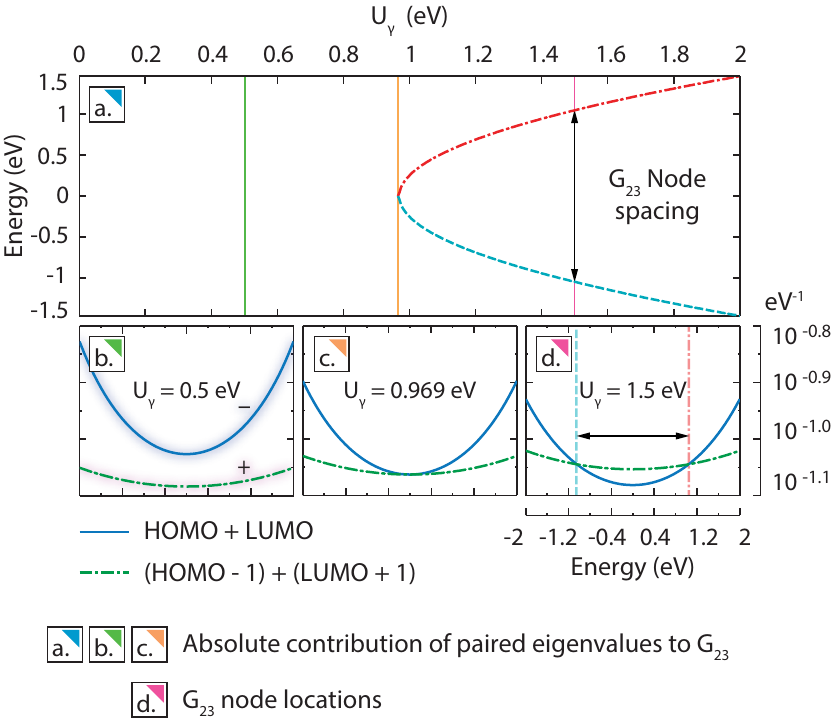}
\caption{The node locations (a) in a four-site extended Hubbard model presented in section \ref{sec:caseStudies} plotted as a function
of an element $U_\gamma$ of the interaction matrix. A degenerate node exists at $E = 0$ for a critical value ($U_\gamma \approx 0.969$) 
but is either split or lifted at other values. In the lifting regime, the eigenvalues of $G$ corresponding to the HOMO
and LUMO make the dominant contribution everywhere within the HOMO-LUMO gap (a-c). In the splitting regime, the dominant
contribution is from the HOMO - 1 and LUMO + 1 eigenvalues in the region bracketed by the nodes (b-d).\label{fourSitesTuning}}
\end{figure}
A curious example of this is furnished by an interacting supernode present in a four-site extended Hubbard model depicted in Figure \ref{fourSitesTuningGeometry},
which is studied in more detail in Appendix \ref{sec:caseStudies}. In the present context the
pertinent observation is that, while this system exhibits a higher-order node at the particle-hole
symmetric point due to series propagation when
there are no interactions, in the presence of Coulomb repulsion this feature is contingent
upon the fine-tuning of an interaction matrix element. When this element deviates from a critical value,
the degenerate node is rapidly split or lifted (Figure \ref{fourSitesTuning} panel a). In
Appendix \ref{sec:caseStudies}, it is shown that the node lifting in this case is attributable to diagrams of the form (e) in
Figure \ref{feynman_diagrams_generic}.
More generally, we find that degenerate nodes generically lie at the boundary between the regimes of splitting and lifting.

Similar behavior can be observed in a noninteractnig system wherein a degenerate node is either split or lifted
depending upon the sign of a perturbation to a hopping matrix-element (Figure \ref{noninteracting_splittingLifting}). 
Indeed, the splitting or lifting of the degenerate node discussed above can be interpreted in topological
terms by considering an effective Hamiltonian defined by $\tilde{H} = H^{(1)} + \Sigma^C_0$. Here $H^{(1)}$ is the noninteracting
part of the extended Hubbard Hamiltonian and $\Sigma^C_0$ is the
Coulomb self-energy evaluated at the node location $E = 0$. In this
simple system, such a Hamiltonian accurately reproduces the Green's function at half-filling
over a wide energy range that includes the HOMO-LUMO gap.

Moreover, in Appendix \ref{sec:caseStudies} it is shown that at the critical tuning associated with the degenerate node
in Figure \ref{fourSitesTuning}, the $(1,4)$
element of this Hamiltonian vanishes due to cancellation between nonlocal exchange and higher order processes.
When this occurs $\tilde{H}$ has the same topology as $H^{(1)}$, consistent with the existence of a degenerate node
despite the presence of nontrivial Coulomb interactions between all the sites.

As a more general alternative to this topological interpretation, it is also possible to understand the crossover between node lifting
and node splitting in terms of electronic structure. We pursue this in the next section, ultimately describing a simple model
that can qualitatively explain the node structure at arbitrary filling in the extended Hubbard systems considered in this work.

\section{Connecting electronic structure with node structure}
\label{sec:electroncStructure}

The node structure of an interacting system can also be understood in terms of electronic structure. To this end, it is useful to consider the eigenvalues
of the retarded Green's function $G$, which are analogous to transmission eigenvalues. In particular, the retarded Green's function of a system with no
coupling to external degrees of freedom can be expressed exactly as:
\begin{equation}
	\label{greensFunctionEigenvalues}
G(E) = \sum_\nu \lambda_\nu |\nu\rangle \langle \nu|
\end{equation}
where $|\nu\rangle$ is an energy-dependent normalized eigenvector of $G$ and $\lambda_\nu$ is the corresponding eigenvalue. 
In Appendix \ref{sec:spectralDecomposition}, it is shown that at zero temperature $\lambda_\nu$ is given exactly by an expression of the form:
\begin{align}
	\label{greensFunctionEigenvalueSimple}
\lambda_{\nu} = \sum_{ \eta } \frac{ Z^\eta_\nu }{E - \Delta E_\eta + i0^+}
\end{align}
Here the index $\eta$ runs over all possible particle- and hole-like transitions out of occupied many-body ground states, and $\Delta E_\eta$ and $Z_\nu^\eta$
are corresponding transition energies and spectral weights.

In principle the number of terms in the sum above is
huge--at least equal to the number of $N \pm 1$ particle many-body
states, where $N$ is the filling of the system. However, absent orbital degeneracy, at energies near the Fermi level 
each eigenvalue is given approximately by an expression of the form:
\begin{align}
	\label{spectralWeightApprox}
\lambda_\nu(E) \approx \frac{\tilde{Z}^\text{p}_\nu}{E - \tilde{\varepsilon}_\nu^\text{p} + i0^+} +  \frac{\tilde{Z}^\text{h}_\nu}{E + \tilde{\varepsilon}_\nu^\text{h} + i0^+} 
\end{align}
where $\tilde{\varepsilon}^{\text{p},\text{h}}$ is an effective parameter that gives the energy cost of creating a particle or hole with state $|\nu\rangle$
and $\tilde{Z}_\nu^{\text{p},\text{h}}$
is a corresponding spectral weight. This expression is developed in Appendix \ref{sec:spectralDecomposition} and the justification for the use of effective parameters 
is considered carefully there. Here it suffices to note that we find this approximation
to be remarkably accurate for the systems considered herein. With long-ranged interactions,
 $\tilde{\varepsilon}^{\text{p},\text{h}}$ can be approximated in terms of noninteracting energies and the charging energy of
the system; similarly, $Z_\nu^{\text{p}} + Z_\nu^{\text{h}} \approx 1$.

\begin{figure}
	\centering
	\includegraphics[scale=1.0]{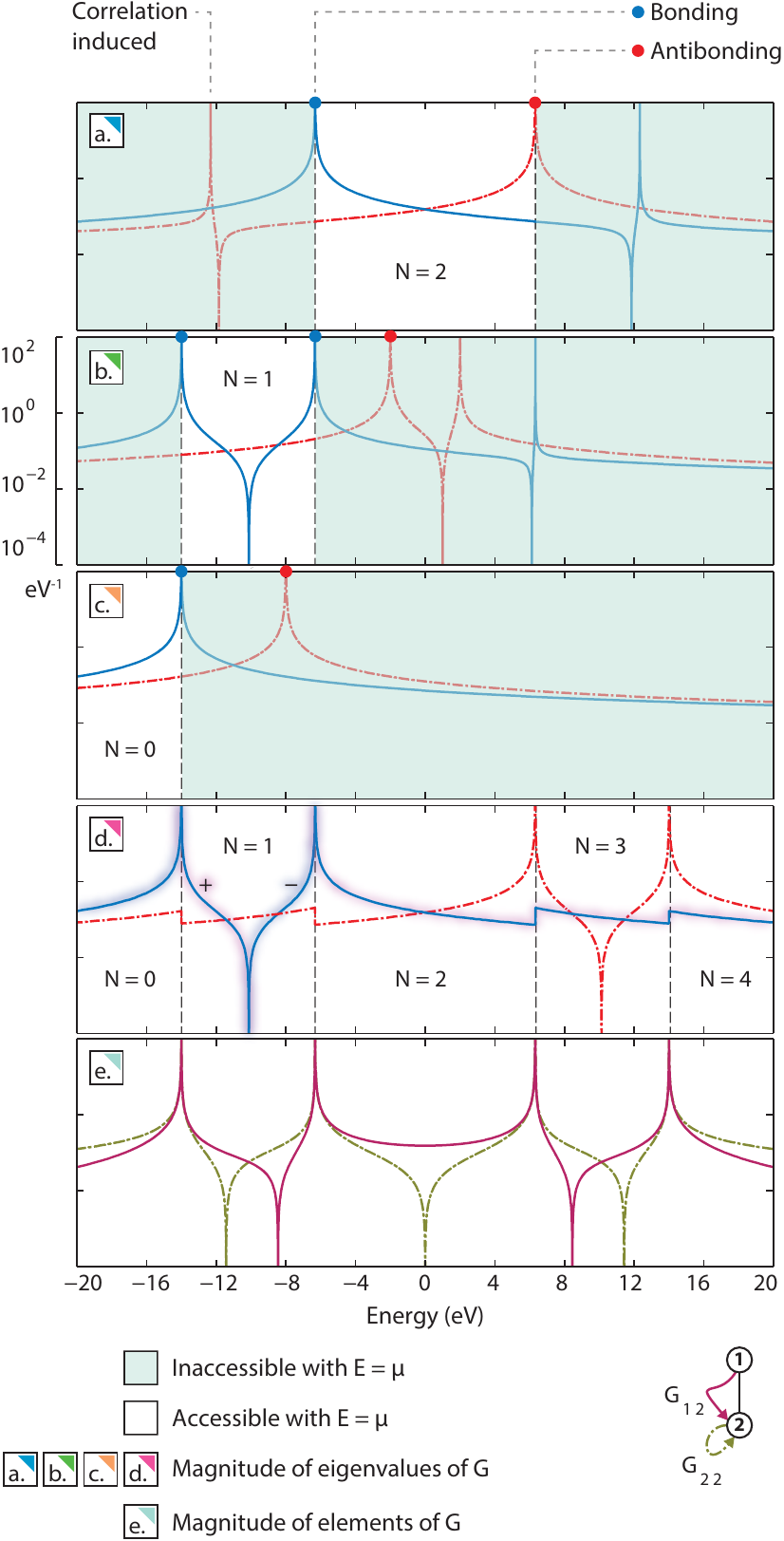}
\caption{Panels (a-d) depict the magnitude of the eigenvalues of the retarded Green's function for a symmetric two-site extended Hubbard
model (system A). In (a), (b), and (c) the system is filled by two, one, and zero electrons respectively, whereas in panel (d) the Green's function is evaluated
at the Fermi level. Panel (e) shows the the matrix-elements $G_{nm}$ of the retarded Green's function, also evaluated at the Fermi level. Nodes in
$G_{nm}$ correspond to intersections of the eigenvalues in panel (d) above.\label{two_sites_matrix_elements}\label{twoSitesGreensFunctionPanel}}
\end{figure}

Together Eq. \eqref{greensFunctionEigenvalues} and Eq. \eqref{spectralWeightApprox} provide an essentially complete qualitative characterization of the low-energy node structure of
a wide range of extended Hubbard models. This is demonstrated by way of example in Figure \ref{twoSitesGreensFunctionPanel}, which shows the exact eigenvalues of a
symmetric two-site extended
Hubbard model from Appendix \ref{sec:caseStudies}.

The approximation expressed by Eq. $\eqref{spectralWeightApprox}$ corresponds primarily to neglecting
the correlation-induced resonances at high energy (those in the shaded region), which are inaccessible to a particle propagating
at the Fermi level. Due to the (anti)symmetry of the eigenvectors corresponding to the bonding and antibonding resonances, the nodes in an element of $G$ occur where
two of the eigenvalues are equal in magnitude. In more general cases the correspondence is modestly less intuitive, but the framework provided by
Eq. $\eqref{spectralWeightApprox}$ still characterizes the node structure accessible at the Fermi level, including the nodes in $G_{nm}$ that occur at odd fillings on either
side of the nodes in the eigenvalues.

In particular, in Appendix \ref{sec:caseStudies} it is shown that the nodes in $G_{nm}$ in
the $N = \pm 1$ filling regions occur due to a node in one of the eigenvalues $\lambda_\nu$, and that their locations are dependent upon the node structure in
the half-filling region. More generally, such nodes $\lambda_\nu$ occur in open-shelled systems. Like the correlation induced
nodes at high energies, they are a signature of non-geometric destructive interference between field configuration histories.

More formally, Eq. \eqref{greensFunctionEigenvalueSimple} implies the energies of nodes in $G_{nm}$ are given exactly by:
\begin{equation}
\sum_{\nu,\eta} \frac{A_\nu^\eta}{E - \Delta E_\eta + i0^+} = 0\\
\end{equation}
where $A_\nu = Z^\eta_\nu \langle n | \nu \rangle \langle \nu | m \rangle$ and $\eta$ ranges
over all transitions between many-body ground states and many-body states with an extra particle or hole.
Putting the terms in this sum above over a common denominator yields the expression:
\begin{equation}
	\sum_{\nu,\eta} A_\nu^\eta \prod_{\gamma \ne \nu}  (E - \Delta E_\gamma) = 0
\end{equation}
where we have neglected infinitesimal imaginary terms.

The order of this polynomial is huge
because it describes the location of all nodes, i.e. even those due to correlations at very high energies. Applying 
Eq. \eqref{spectralWeightApprox} renders it more intuitively useful, in which case we have:
\begin{equation}
		\label{nodePolynomial}
\sum_{\nu \atop \{\text{p},\text{h}\}} \tilde{A}_\nu^{\text{p},\text{h}}\prod_{\mu \ne \nu}  (E - \tilde{\varepsilon}_\eta^{\text{p},\text{h}}) = 0
\end{equation}
where both $\mu$ and $\nu$ range over single-particle eigenvectors. The degree of the polynomial
is then at most $2(N - 1)$, where $N$ is the number of single-particle eigenstates. If we
take into account the considerations in Appendix \ref{sec:spectralDecomposition}, it can be seen that
a polynomial of order $N - 1$ suffices when the system is in a closed-shelled state. In practice,
only a few of these roots are present at low-energies. Put another way, the number of nodes
at low energy is at most half the number of eigenvalues competing in this region.

In light of the present discussion, the node structure depicted in Figure \ref{fourSitesTuning}, explained topologically in the preceding section, can
also be understood either in terms of the interference of the eigenvalues of $G$ (panels a-c) or in terms of the roots of the polynomial
in Eq. \eqref{nodePolynomial}. In the former case the crossover between lifting and splitting occurs as the pair
of eigenvalues making the dominant contribution to $G$ near $E = 0$ switches from the HOMO and LUMO to the HOMO - 1 and
the LUMO + 1. In the latter view, lifting occurs when Eq. \eqref{nodePolynomial} has complex roots whereas
the splitting regime corresponds to real roots.

These equivalent perspectives generalize readily to the case of more complex systems, and imply that the node structure in similar
models satisfies a few generic conditions: (1) \emph{there is a strong tendency toward the preservation of the parity (i.e. evenness or oddness) of the number of nodes}; (2) \emph{all degenerate
nodes lie at the boundary between the regimes of node splitting and node lifting}; (3) \emph{nodes in the eigenvalues
of the Green's function occur in open-shelled states and at high energies due to non-geometric interference between field configuration histories
and are responsible for the Mott nodes in $G_{nm}$}; (4) \emph{the sensitivity
of a node to perturbation or the tuning of a parameter increases with its degeneracy}; (5) \emph{the tendency of degenerate
nodes to lift rather than split increases with their degeneracy}.

The last two points can be understood heuristically from the geometry of the eigenvalues $\lambda_\nu$. Alternatively,
they can inferred formally based on Eq. \eqref{nodePolynomial} and 
well-known work\cite{wilkinsonPaper} regarding the response of polynomial roots to the perturbation of their coefficients. In particular,
the degenerate roots of a polynomial are an ill-conditioned function of its coefficients. The same work implies that 
even the location of nondegenerate nodes may in principle be an ill-conditioned function of parameters in the Hamiltonian, although this
is expected to be a rare case.

These properties are exemplified by the
extreme sensitivity of degenerate geometric nodes to any perturbation inconsistent with series
propagation. For example,
for two systems arranged in series that do not interact with each other but are themselves
interacting, we find in Appendix \ref{sec:caseStudies} that processes fundamentally
inconsistent with the notion of series propagation are sufficient to dramatically destroy degenerate
nodes. Remarkably, in the same models these small deviations from pure series propagation
have little effect on other low-energy features.

\section{Conclusions}

The retarded Green's function of an interacting system has the potential to exhibit rich and nontrivial node structure.
However, in this work we have made a few generic observations:
In extended Hubbard models (1) there is a strong tendency toward the preservation
of the parity (i.e. evenness or oddness) of the number of nodes; (2) degenerate nodes require fine-tuning in the presence
of interactions and sit at the boundary between regimes of node splitting and node lifting;
(3) nodes in the eigenvalues
of the Green's function occur in open-shelled states and at high energies due to non-geometric interference between field configuration histories; 
(4) the sensitivity of a degenerate node to the tuning of a parameter increases with its degeneracy;
and (5) the tendency of degenerate nodes to lift upon perturbation rather than split increases with their degeneracy.

These properties can be understood in terms of electronic structure by way of a simple approximation that qualitatively explains the node structure
of the extended Hubbard models we considered, regardless of filling. More formally, nodes of the retarded Green's function
are the roots of a polynomial. In some cases, node structure can also be understood
in topological terms. However, in the presence of interactions, we find that geometric degenerate nodes predicted to exist on the
basis of series propagation are present if and only if a definition of series propagation formulated herein is imposed artificially. This is true even when there are no
interactions \emph{between} the units arranged in series.

The reason for this is fundamental: Amplitudes in interacting systems are not expressible as a sum of \emph{geometric} Feynman paths. All possible field
configuration histories contribute
coherently to an interacting propagator. Among the low energy interference features in interactions systems, degenerate nodes appear to be uniquely sensitive to this.

\begin{acknowledgments}
This material is based upon work supported by the Department of Energy under Award Number DE-SC0006699.
\end{acknowledgments}

\appendix
\section{Case studies: Detailed analyses of the node structure in select extended Hubbard models}
\label{sec:caseStudies}

\begin{figure}
	\centering
	\includegraphics[scale=1.0]{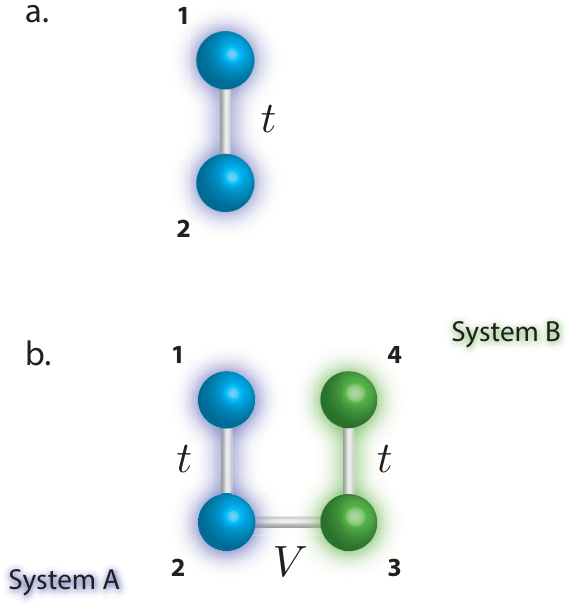}
\caption{Top (a): Two sites in an extended Hubbard model linked by a matrix element $t$. Bottom (b): Two systems of the form in (a) linked
by a matrix element $V$ and possibly with some Coulomb interactions between them. This is a realization of Figure \ref{series_transport_boxes} if we identify sites one and
two as system A and sites three and four as system B.\label{two_sites} \label{four_sites}}
\end{figure}
In the following we present detailed analyses of the node structure in select two and four site extended Hubbard models.
We also touch briefly upon the node structure in larger systems. The material presented herein forms the basis
for many of the observations made earlier in this work, but is necessarily more detailed and more technical
than what has come before this point. The reader is encouraged to study the portions of interest to him or her.

Our starting point is Figure \ref{four_sites}, which depicts a
pair of two-site systems, $A$ and $B$, joined by a single hopping matrix element $V$. The interactions are of the
extended Hubbard form so that the full Hamiltonian is:
\begin{equation}
\mathcal{H}^{A + B} = \sum_{n, m, \sigma}H^{(1)}_{nm} d^\dagger_{n\sigma} d_{m\sigma} + \frac{1}{2}\sum_{nm} U_{nm} \rho_n\rho_m
\end{equation}
where $\rho_n = \sum_{\sigma} d^\dagger_n d_n - 1$, which ensures particle-hole symmetry, and:
\begin{equation}
	H^{(1)}_{nm} = \begin{pmat}({|})
\mathbf{H}_A & \mathbf{H}_{AB} \cr \-
\mathbf{H}_{AB}^\dagger & \mathbf{H}_B \cr
\end{pmat} =
\begin{pmat}({.|.})
\epsilon & -t & 0 & 0 \cr
-t^* & \epsilon & -V & 0 \cr \-
0 & -V^* & \epsilon & -t \cr
0 & 0 & -t^* & \epsilon\cr
\end{pmat}\nonumber
\end{equation} 
\begin{equation}
	U_{nm} = \begin{pmat}({|})
\mathbf{U}_A & \mathbf{U}_{AB} \cr \-
\mathbf{U}_{AB}^\dagger & \mathbf{U}_B \cr
\end{pmat} =
\begin{pmat}({.|.})
U & U_\alpha & U_\beta & U_\gamma \cr
U_\alpha & U & U_\alpha& U_\beta \cr \-
U_\beta & U_\alpha & U & U_\alpha \cr
U_\gamma & U_\beta & U_\alpha & U\cr
\end{pmat} \nonumber
\end{equation}

This system has been studied as a model of a cross-conjugated molecule elsewhere,\cite{justinMottNodePaper} but here we do not concern ourselves with
detailed models of electronic structure that might obscure conceptual issues. 
Instead, we take a deliberately simple parameterization and tinker with it.
In particular, we begin with:
\begin{gather}
\epsilon = 0 \; \text{eV}, \quad t = 3 \; \text{eV}, \quad U = 10 \; \text{eV}, \quad U_\alpha = 6 \; \text{eV}, \nonumber \\
 \quad U_\beta = 4 \; \text{eV}, \quad V = t, \quad U_\gamma = U_\alpha \label{initialParameterization}
\end{gather}
This parameterization is roughly consistent
with interactions that might be expected on the molecular scale,\cite{barrPiEFTPaper,ohnoPaper,castletonPaper} in particular for systems with a geometry
respecting the symmetries of (b) in Figure \ref{four_sites}. 

\subsection{Two sites (System A)}

We first consider system $A$ in isolation. The elements $G_{nm}$ of the retarded Green's function in this case 
are plotted in Figure \ref{two_sites_matrix_elements} panel (e) in the zero-temperature\footnote{Room temperature calculations were carried out and yielded results visually indistinguishable from those presented
here.} limit but with the system maintained at a fixed chemical potential $\mu$ set equal to the energy $E$. Here,
 $\mu$ simply dictates the filling of the system. Panels (a-d) show the eigenvalues of the retarded Green's function either at fixed filling
(a-c) or with the chemical potential equal to the energy (d). The shaded blue areas are the regions that are inaccessible to particles
propagating at the Fermi level, i.e. the regions that do not contribute to linear response transport.\cite{jauhoBook, dattaBook}

We now comment upon several noteworthy features in this figure:

\paragraph{The node in $G_{22}$ at $E = 0$ eV:}
The noninteracting Green's function (not shown) exhibits a node at half-filling at $E = 0$. Formally, this is
an antiresonance that occurs at the on-site potential of
site one regardless of the matrix element $t$. Alternatively, it manifests as destructive interference between propagation mediated
by the bonding and antibonding orbitals,
i.e. the eigenstates of the Green's function with positive and negative parity respectively.

In panel (e) it is evident that this node persists in the interacting Green's function $G_{22}$. In fact, there is a simple
symmetry argument that requires this: While in the interacting case there are no
single-particle orbitals, the eigenvectors of $G$ are still dictated entirely by symmetry and therefore identical
to their noninteracting counterparts. In the present case, this observation is sufficient to imply that the node in $g_{22}$ must also exist in $G_{22}$.
Alternatively, this could be viewed as a consequence of the Luttinger theorem.\cite{luttingerTheorem1,luttingerTheorem2,justinMottNodePaper}

\paragraph{Singularities in the bonding and antibonding eigenvalues:}
In Appendix \ref{sec:spectralDecomposition}, the eigenvalues $\lambda_\nu$ are given by the coherent sum of amplitudes associated
with processes wherein a particle or hole is
added to a many-body ground state.
Absent interactions, each of these processes is simply the addition of a particle or hole to a noninteracting orbital,
and the energy cost for corresponding particle-like and hole-like processes differs only in sign. Under these
circumstances, the resonances for both kinds of processes occur at the same energy and there is one singularity per
eigenvalue. With interactions this breaks down for two reasons: (1) When the system is charged (here, away from half filling)
the energy cost of adding a particle to the system is no longer equal in magnitude to that of a hole; (2)
transitions between ground states and correlated excited states lead to narrow resonances at high energies.

Reason (1) gives rise to Coulomb blockade\cite{averinCoulombBlockadeTheory,fultonCoulombBlockade,giaeverNascentCoulombBlockade,giaeverNascentCoulombBlockade2} in the context of transport,
and occurs generically in presence of repulsive long-ranged interactions.
It can be examined separately from (2) by taking $U_{nm} = U$, in which case the repulsive Coulomb energy depends only
on the net charge of the system and the eigenvectors $\{|\nu\rangle\}$ are exactly equal to the noninteracting ones.
In either cases (1) or (2), each singularity corresponds to some physical process wherein a particle or hole is added to the system
in a manner that respects the symmetry of the corresponding eigenvector. This leads to novel node structure, as described below.

\paragraph{The node in the bonding eigenvalue at $N = 1$ near $E = -10$ eV:}
When $N = 1$ the bonding orbital is half-filled and thus accommodates both particle-like and hole-like addition. As noted above,
in the presence of Coulomb interactions the resonances associated with these processes occur at distinct energies. Midway between
them ($E \approx -10$ eV)
the energy needed to create a virtual particle
in this orbital is equal to that of a virtual hole, and so these processes contribute opposite amplitudes, completely
suppressing propagation mediated by this orbital.

Note that the crucial fact here is not so much that the interference
is between particle-like and hole-like propagation--this is formally responsible for nodes in $G$ including the antiresonance in $G_{22}$--but
that the destructive interference is between two \emph{processes} with different amplitudes that are mediated by the
\emph{same} orbital.\footnote{More precisely, associated with the same eigenvalue of $G$; the notion of a single-particle orbital here is formally justifiable when $U_{nm}$ is constant.}
This is totally alien to
noninteracting systems and hence is beyond the scope of \emph{geometric}
Feynman paths, instead arising from the interference of field configuration histories. Nodes of this form are actually a
general phenomenon that can be seen to arise between other singularities in the eigenvalues of $G$. Such nodes correspond to the coherent
interference of histories (processes) involving states with the same symmetry. In a noninteracting system, they do not exist.

\paragraph{The nodes in $G_{nm}$ at $N = 1$ near $E = -10$ eV:}
The aforementioned nodes in the eigenvalues of $G$ are associated with nodes that occur in the elements of $G$ around
the same energy. Related nodes have been reported previously in the theoretical transmission spectrum
of molecular radicals\cite{justinMottNodePaper} and in the Hubbard model.\cite{mottNodePaper} In the
present context, a node in the bonding eigenvalue can be seen around $-10$ eV in the $N = 1$ region.
At this energy the positive parity eigenvector makes no
contribution to the Green's function and $G$ is perfectly antisymmetric, i.e. $G_{11} = -G_{12}$.  This feature is
bracketed by nodes in $G_{11}$ and $G_{12}$ that occur due to destructive interference between propagation through the
bonding and antibonding resonances, i.e. at the locations where the magnitude of the eigenvalues intersect in panel (d). Between these nodes,
propagation is mediated almost entirely by the antibonding
resonance, despite the proximity of the bonding resonances.

Considering energies from left to right in the figure,
 as the bonding resonance is suppressed all the nodes
in $G_{nm}$ in the half-filling region reappear to the left of the node in the eigenvalue around $-10$ eV. As the
bonding resonance opens back up, one node appears for each element of $G$ that does \emph{not} exhibit a node in the
half-filling region. Thus the location of a node in $G_{nm}$ at $N = 1$ is related to the existence or nonexistence
of a node in the same element of $G_{nm}$ at half-filling. An exceptional case is when the suppressed eigenvector is
the only one that contributes to a particular element of the Green's function; in this situation the node in that element
of $G$ coincides with the node in the eigenvalue. Till now this is the only case that has been studied, but 
here we point out that it is
the exception rather than the rule. Overall, this behavior generalizes to larger systems, although the structure of
the eigenvalues there is richer and can be complicated by, e.g., avoided crossings.

We now consider in a similar manner the node structure associated with the composite system $A + B$.

\begin{figure}
	\centering
	\includegraphics[scale=1.0]{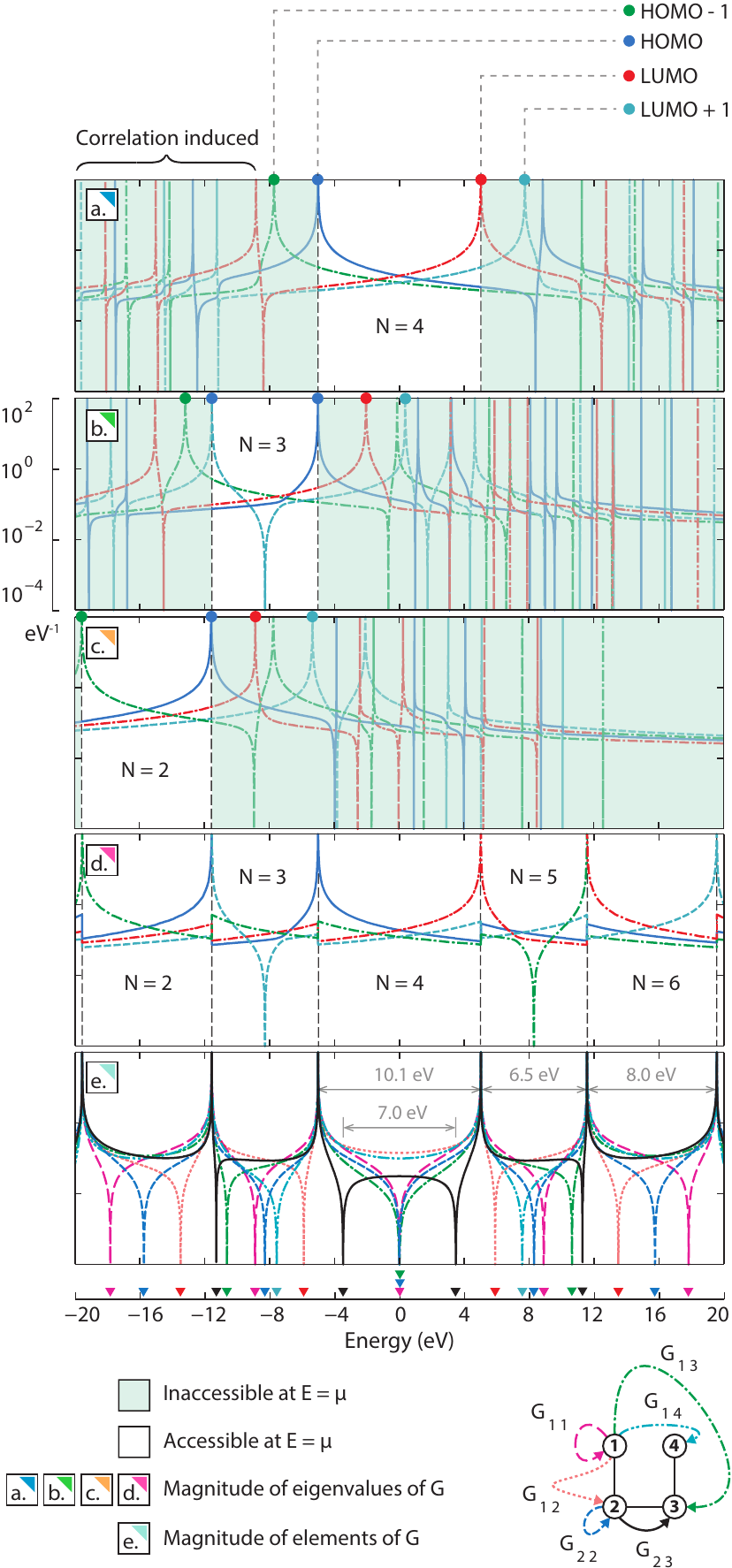}
\caption{Panels (a-d) depict the magnitude of the eigenvalues of the Green's
function for system $A + B$ in a manner analogous to
Figure \ref{twoSitesGreensFunctionPanel}. The associated eigenvectors (not shown) are now energy-dependent, but can still be classified based on symmetry.  Panels (e) and (f) show the elements of the interacting
and noninteracting Green's functions $G_{nm}$. The $G_{23}$ element exhibits a pair of nodes that arise from the splitting of a degenerate node present in the absence of interactions.
In the odd-filling regions a node in an eigenvalue of $G$ causes clusters of nodes in $G_{nm}$.}
\label{fourSitesGreensFunction} \label{fourSitesGreensFunctionNoninteracting}
\label{fourSitesGreensFunctionPanel}
\end{figure}

\subsection{Four sites (System A + B)}

We now consider the composite system depicted in Figure \ref{four_sites} (b) wherein system $A$ is coupled to an identical system $B$
by a matrix element $V$, forming a composite system $A + B$.  The node spectrum in this case is depicted in Figure \ref{fourSitesGreensFunction} in a manner
analogous to Figure \ref{two_sites_matrix_elements}, which depicts the node structure of system $A$ alone
and is discussed at length in the foregoing section.

Most of the observations
made there carry over to the case of the four-site system under consideration here. For example, the
node structure at odd filling is related in the same way to nodes in the eigenvalues of the Green's function. 
However, for the remainder of this section we focus our attention on the case of half-filling. In particular,
the node structure of $G_{23}$ is noteworthy here:

Without interactions (not shown) the Green's function of system $A + B$ is given by Eq. \eqref{pureSeriesEnergyDomain}.
In this case, the node at $E = 0$ in $G_{22}$ in system $A$ and its counterpart in system $B$ give rise to a degenerate
node at $E = 0$ in system $A + B$ where $G_{23} \propto E^2$. However, as reported previously,\cite{justinNodeSplittingPaper}
in the presence of interactions this is not the case. Instead, the degenerate node is split into two ordinary nodes which appear around $\pm 4$ eV
in panel (e) for the parameterization given by Eq. \eqref{initialParameterization}. Since this is contrary to the predictions of Eq.
\eqref{pureSeriesEnergyDomain}, this node splitting is inconsistent with the definition of series propagation discussed
earlier. It is thus attributable to some amalgam of the diagrams in Figure \ref{feynman_diagrams_generic}.

To shed light on what is happening, we use exact diagonalization to isolate several interesting combinations of these diagrams,\footnote{To
be precise, nonperturbative effects are included as well; therefore, in this context the diagrams mentioned should be viewed as a conceptual device used
to classify processes. The only calculation performed diagrammatically was self-consistent Hartree-Fock.}
as well as consider a variety of qualitatively interesting variations on the parameterization
given by Eq. \eqref{initialParameterization}. Along the way, we demonstrate results highlighted earlier
in this work via a careful consideration of these cases. In particular, we show that the node splitting phenomenon just discussed
occurs because degenerate nodes are extremely sensitive to the breakdown of series propagation, even when
there are no interactions between systems $A$ and $B$.

The cases considered are organized into the panels in Figure \ref{fourSitesTheoryComparison}, which we now remark upon individually:

\begin{figure*}
\centering
\includegraphics[scale=1.0]{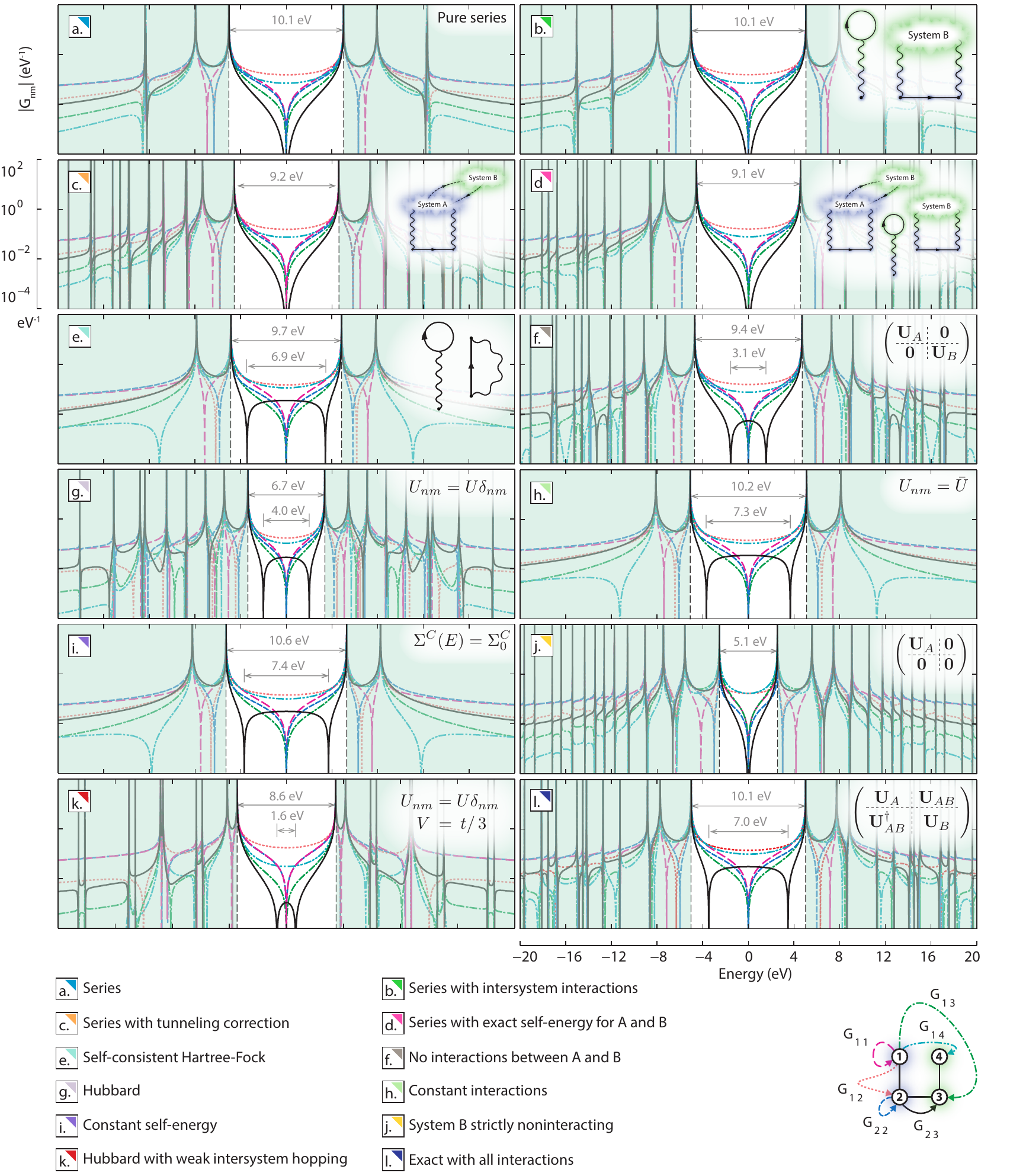}
\caption{The absolute elements of the retarded Green's function of system $A + B$
at half-filling, calculated via exact diagonalization. The calculation is performed with various parameterizations or in such a manner as to selectively
include the processes depicted schematically in Figure \ref{feynman_diagrams_generic}. 
In this case, the presence of diagrams of the form (d) or (e) is necessary and sufficient to destroy the degenerate node
in $G_{23}$.\label{fourSitesTheoryComparison}
Other features in the low-energy region accessible to particles at the Fermi level (white) show no such sensitivity here.}\end{figure*}

\paragraph{Node structure with no diagrams from Figure \ref{feynman_diagrams_generic}:}
By calculating the exact Green's functions of systems $A$ and $B$ in isolation and inserting these into
Eq. \eqref{pureSeriesEnergyDomain}, we may calculate the Green's function of system $A + B$ excluding
precisely the diagrams depicted in Figure \ref{feynman_diagrams_generic}. This is equivalent
to enforcing the stritest definition of series propagation proposed in section \ref{sec:seriesPropagation}, i.e. one wherein
$G^A$ and $G^B$ in Eq. \eqref{pureSeriesEnergyDomain} are not dressed by the coupling of system $A$ to system $B$.

The Green's function of $A + B$ calculated
in this manner is shown in panel (a), and a degenerate
node can be seen in $G_{23}$. This occurs because the elements $G^A_{22}$ and $G^B_{33}$
of the isolated Green's functions of $A$ and $B$ (not shown) exhibit nodes at $E = 0$, and,
with series propagation enforced, these necessarily give rise to a degenerate node in the
composite system.

We now probe the effects of including one or more diagrams that are inconsistent with
at least one of the definitions of series propagation considered earlier.

\paragraph{Node structure with diagrams of the form (a) + (b):}
A priori, (a) and (b) in Figure \ref{feynman_diagrams_generic} are plausible candidates for the
splitting of the degenerate node in $G_{23}$, and we can include these diagrams
selectively via exact diagonalization. For combinations of (a) and (b) this entails calculating $G^A$
and $G^B$ including intersystem Coulomb interactions but with $V = 0$. Inserting these into Eq.
\eqref{pureSeriesEnergyDomain} then gives the
desired Green's function, which is consistent with a weak definition of series propagation wherein $G^{A,B}$ are
dressed to account for correlations caused by long-ranged Coulomb interactions between $A$
and $B$.

The Green's functions $G^{A,B}$ calculated during this procedure (not shown) still exhibit nodes at $E = 0$.
Consequently, the the Green's function of $A + B$ calculated in this manner, which
is depicted in panel (a), has a second order degenerate node. Thus,
correlations induced by long-ranged Coulomb interactions between $A$ and $B$ are not responsible for splitting the
degenerate node in this case.

\paragraph{Node structure with diagrams of the form (c):}
Diagrams of the form (c) can also be included selectively
using exact diagonalization. To this end, the full Green's function of
$A + B$ is first calculated without long-ranged Coulomb
interactions but with $V \ne 0$. The corresponding Coulomb self-energy
is then determined using the Dyson equation, and the parts of this associated with self-energies
for systems $A$ and $B$ are extracted. These are used to determine the Green's functions
$G^{A,B}$ dressed by the diagram shown, which are in turn used in Eq. \eqref{pureSeriesEnergyDomain}.

The elements $G_{22}^{A}$ and $G_{33}^{B}$ of the dressed Green's functions of systems $A$ and $B$ that
are obtained during this procedure (not shown) still
exhibit nodes at $E = 0$. Consequently, the degenerate node in the Green's function of system $A + B$ persists.

\paragraph{Node structure with diagrams of the form (a) - (c)}
Combinations of diagrams (a)-(c) may be selected for via exact diagonalization using almost
the exact same procedure as for diagrams (c) alone. The only change necessary is
to include long-range Coulomb interactions in the initial exact calculation of the Green's
function of system $A + B$.

Again, the dressed Green's functions of $A$ and $B$ (not shown) possess nodes at $E = 0$ that cause
degenerate nodes in the Green's function for system $A + B$ when it is calculated using
Eq. \eqref{pureSeriesEnergyDomain}.

\paragraph{Node structure with diagrams of the form (a) + (d) (self-consistent Hartree-Fock):}
To probe the effects of diagrams of the form (a) and (d), self-consistent Hartree-Fock calculations
were performed to determine the Green's function of system $A + B$. The degenerate
node is split, and calculations
including only the Hartree diagram (not shown) indicate that the Fock diagram is responsible.

Thus nonlocal exchange is \emph{sufficient} to split the degenerate node; however,
it is not necessary or even the dominant contribution to the node splitting, as the remaining
cases demonstrate.

\paragraph{No interactions between systems A and B:}
Here the Green's function of $A + B$ is calculated
exactly with no interactions between $A$ and $B$ (i.e. $\mathbf{U}_{AB} = \mathbf{0}$). The degenerate
node is split by roughly $3$ eV and, on the basis of the foregoing cases and the
fact that there are no intersystm interactions whatsoever, we conclude
that diagrams of the form (e) are responsible. 

\paragraph{Hubbard interactions:} Here the Green's function of $A + B$ is calculated
exactly with Hubbard interactions (i.e. $U_{nm} = \delta_{nm} U$). Again, a very
large splitting is present despite no long-ranged interactions whatsoever. In light
of the other cases, we conclude diagrams of the form (e) are responsible.

\paragraph{Constant interactions:}  As it is mentioned repeatedly in this work,
we consider here the case of constant interactions,
i.e. $U_{nm} = \bar{U}$ where $\bar{U}$ is the average of the full
interaction matrix from Eq. \eqref{initialParameterization}. Again a large splitting is present in $G_{23}$.

\paragraph{Fixed Coulomb self-energy:} As it is relevant elsewhere in this work,
we present here the case wherein the Coulomb self-energy $\Sigma^C$ is fixed to its value at $E = 0$.
The splitting shown is attributable to the elements
of the Coulomb self-energy that connect systems $A$ and $B$. These are precisely the elements that arise from
diagrams of the form (d) and (e), or amalgams thereof.

\paragraph{System B strictly noninteracting:} Here system $B$ is strictly noninteracting. Diagrams
of the form (d) and (e) are not present in this scenario, and consistent with this the degenerate node survives.

\paragraph{Hubbard interactions with weak intersystem hopping:} Here the Green's function for system $A + B$ is calculated with
 Hubbard interactions $U_{nm} = \delta_{nm} U$ and
$V = t/3$ rather than $V = t$ as in the other cases. This scenario has the weakest overall coupling
between the systems, and consistent with this, the smallest splitting of the degenerate node. However, the 
node splitting is still significant.

\paragraph*{Figure \ref{fourSitesTuning}. A degenerate node from fine-tuned interactions:}
Finally, we present an example of a degenerate node that exists within an interacting system despite the presence
of diagrams that are inconsistent with series propagation. The cost is that instead of its existence being
ensured by the geometry of the system, the degenerate node requires fine-tuning of, for example, an interaction matrix
element. Moreover, this example also demonstrates that a perturbation can destroy a degenerate node by 
of lifting rather than splitting.

We consider the system $A + B$ with the parameterization given by $U = U_\alpha = U_\beta = 5$ eV, $U_\gamma = 0$ eV.
In this case the supernode present within noninteracting models is not split,
but completely lifted. This persists as $U_{14} = U_\gamma$ is increased, up to a critical value of $\approx 1$ eV near
which there is a sharp transition between the regimes of node lifting and splitting.
At the critical value a supernode exists despite the presence of interactions that are inconsistent
with series propagation, but at the cost of fine-tuning.

\begin{figure}
	\centering
	\includegraphics[scale=1.0]{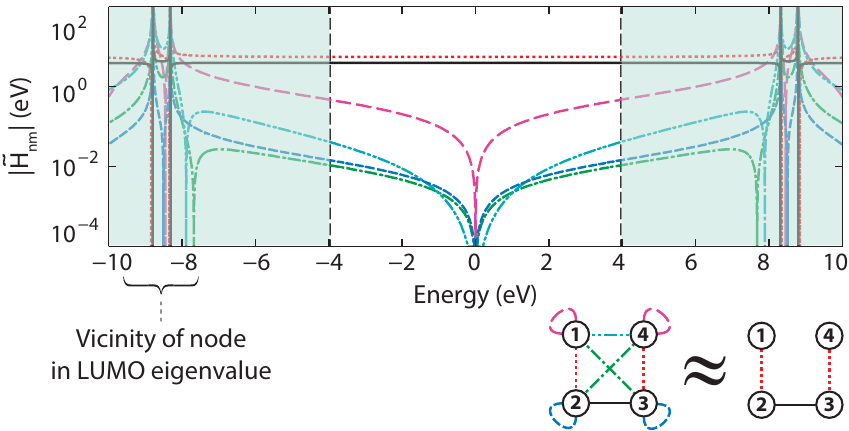}
\caption{The absolute matrix elements of the family of single-particle Hamiltonians defined by Eq. \eqref{onebodyMapping}. The
parameters of the four-site extended Hubbard model are tuned such that $\tilde{H}_{14} \approx 0$ near $E = 0$. Since $\tilde{H}_{11}$,
$\tilde{H}_{22}$, and $\tilde{H}_{13}$ are suppressed at this point by particle-hole symmetry, the system
has a ``low-energy topology'' that exhibits a degenerate node in the element $G_{23}$ of the associated Green's function. This
approximation breaks down near nodes in the eigenvalues of the Green's function (singularities in the self-energy), as can be seen around $\pm 8$ eV.}
\label{fourSitesEffectiveOnebody}
\end{figure}
To investigate this phenomenon, it is instructive to use the exact Coulomb self-energy $\Sigma^C(E)$ to define a formal
mapping of the many-body system $A + B$ onto a family of single-particle systems with Hamiltonians:
\begin{equation}
	\label{onebodyMapping}
\tilde{H} = H_{A+B}^{(1)} + \Sigma^C(E)
\end{equation}
where  $H_{A+B}^{(1)}$ is the noninteracting portion of the Hamiltonian of system $A + B$. This reproduces the exact one-body Green's function by construction.
The elements of $\tilde{H}$ are plotted in Figure \eqref{fourSitesEffectiveOnebody} for the critical value $U_\gamma = 0.969$ eV.
Regardless of tuning, elements  corresponding to next-next nearest neighbor hopping (e.g. $\Sigma^C_{13}$) break particle-hole symmetry and
are therefore suppressed at half-filling near the particle-hole
symmetric point ($E = 0$). At the critical value of $U_\gamma$, nonlocal exchange cancels exactly with higher order diagrams and
the element $\Sigma^C_{13})$ also vanishes, whereas for larger (smaller) values of $U_\gamma$ it is respectively of the same or opposite sign
as the hopping matrix elements in $H_{A+B}^{(1)}$. Thus, in this case, the competition between diagrams (d) and (e) leads to the switchover between node lifting
and splitting depicted in Figure \ref{fourSitesTuning}. Conclusive proof of the important qualitative role of higher order
processes is furnished by self-consistent Hartree-Fock calculations (not shown), which do not lead to node lifting for this model.

As an aside, we note here that the approximate Hamiltonian obtained by setting $E = 0$ in the exact expression \eqref{onebodyMapping} is valid only away
from nodes in the eigenvalues of the Green's functions (singularities in the Coulomb self-energy).

\begin{figure}
	\centering
	\includegraphics[scale=0.95]{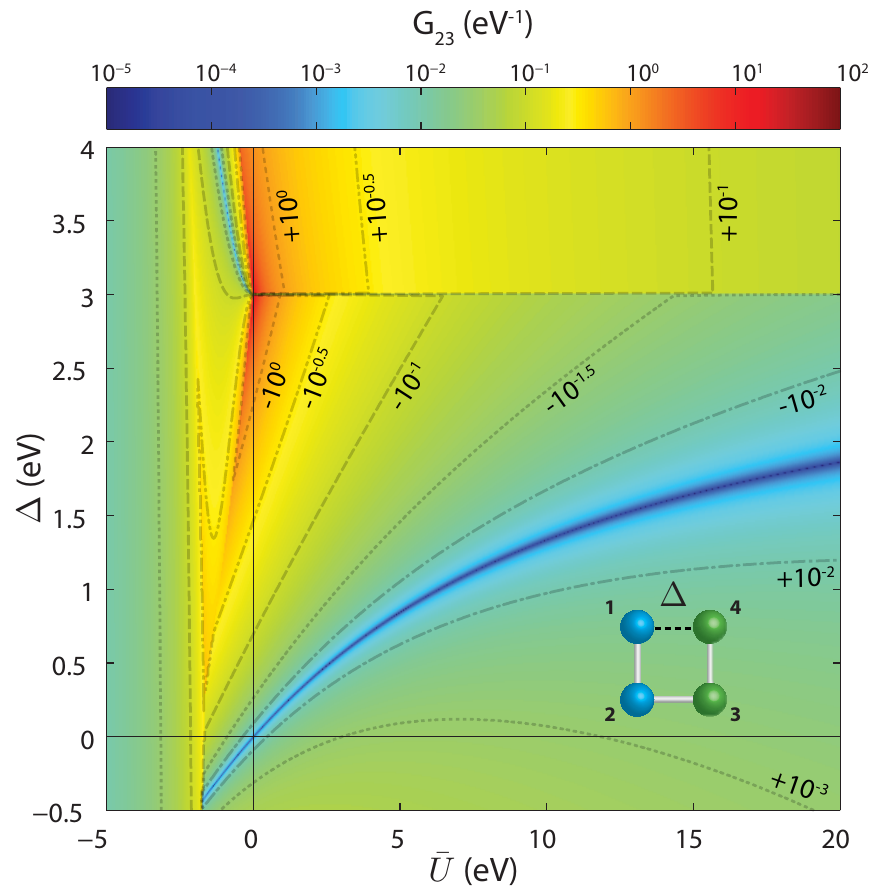}
\caption{The element $G_{23}$ of the Green's function for system $A + B$ evaluated at the particle-hole symmetric point, plotted as a function of the average interaction matrix element $\bar{U}$
and a hopping matrix element $\Delta$ connecting sites one and four. For simplicity, we have taken $V = t$ and constant interactions, i.e. $U_{nm} = \bar{U}$. A degenerate node exists
for a family of parameterizations corresponding
to the arc-shaped path that intersects $\bar{U} = 0$ eV, $\Delta = 0$ eV. However, it is forbidden by symmetry in the special case of $\Delta = t$. Absent the exact symmetry present
in this case, the 
interplay between topology and interactions is complex.\label{fourSites_nodeVsUandT14}}
\end{figure}

\paragraph*{Figure \ref{fourSites_nodeVsUandT14}. A degenerate node from fine-tuning between a hopping matrix element and a charging energy $\bar{U}$ with
constant interactions:}
The generic dependence of degenerate nodes upon fine-tuning in the presence of interactions can be seen by considering the amplitude
$G_{23}(E)$ evaluated at $E = 0$ as a function of parameters in the Hamiltonian. For simplicity, this is done in Figure \ref{fourSites_nodeVsUandT14} for system $A + B$ with all the interaction matrix elements
set to a constant $\bar{U}$, with $V = t$, and a with a hopping matrix element $\Delta$ introduced between sites one and four. A degenerate supernode traces a path that corresponds to a family of models wherein
$\tilde{H}_{14} = 0$. Just below this region the node is split, and above it, lifted.

The case $\Delta = t$
is also special in that symmetry then precludes the existence of a node at $E = 0$ regardless of the charging energy $\bar{U}$. This
can be argued formally based on the eigenvectors of the Green's function, which here are just equal to the noninteracting ones. More generally,
the same claim holds for non-constant interactions because in the presence of such symmetry, the interacting eigenvectors can be taken as equal to the noninteracting ones.

\begin{figure}
	\centering
	\includegraphics[scale=0.95]{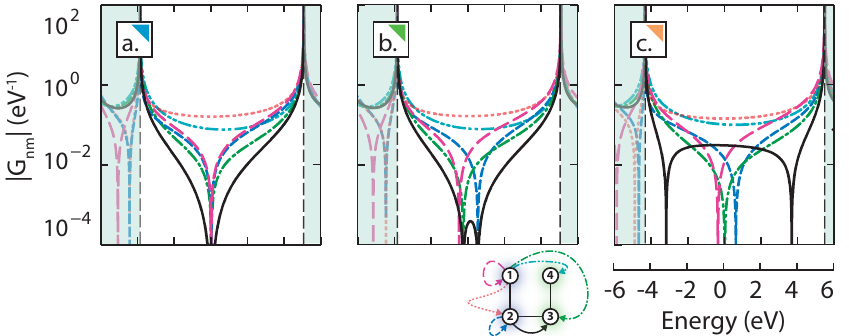}
\caption{The absolute elements of the retarded Green's function of a variation on the extended Hubbard system $A + B$, calculated at half-filling via exact diagonalization. Here
the on-site energy of site $3$ in system $B$ is shifted up by $3$ eV so that systems $A$ and $B$ are not identical and, moreover, there is no particle-hole symmetry.
Panel (a) shows the results for pure series propagation (no diagrams from Figure \ref{feynman_diagrams_generic}), panel (b) shows the results with the inclusion of diagrams (a)-(c), and panel (c) shows the exact solution.\label{brokenSymmetryPanel}}
\end{figure}

\paragraph*{Figure \ref{brokenSymmetryPanel}. Broken symmetry between $A$ and $B$:}
Up to this point we have only considered cases wherein systems $A$ and $B$ are identical. If this is not so, then it should perhaps not be suprising that
a degenerate node expected on the basis of the alignment of nodes in $A$ and $B$ is disrupted. Figure \ref{brokenSymmetryPanel} explores this,
depicting the Green's function when the on-site potential of site $3$ is shifted up by $3$ eV:
 
Here panel (a)
has only diagrams of the form (c), which, as before, do not perturb the node structure in $G^{A,B}$ (not shown) and therefore do not disrupt the nodes in $G^{A + B}$.
Panel (b) has all diagrams of the form (a)-(c) and collectively these
do split the degenerate node now. Since series propagation is enforced here, this must occur via a disruption of the relevant nodes in $G^{A,B}$.
Moreover, to cause splitting these nodes must be shifted by different amounts, which requires that $A$ and $B$ be distinct. In this
case it also necessitates broken particle-hole symmetry so that the nodes are not pinned at $E = 0$.

It is interesting to note, however, that diagrams (a)-(c) still only make a small contribution to the overall node splitting, as can be seen in panel (c),
which shows the exact solution. Thus, at least in this case, the disruption of the degenerate node is not due primarily to changes in the electronic
structure of $A$ or $B$ as defined by the dressing of $G^{A,B}$.

\subsection{Greater than four sites}
\begin{figure}[h]
	\centering
	\includegraphics[scale=0.95]{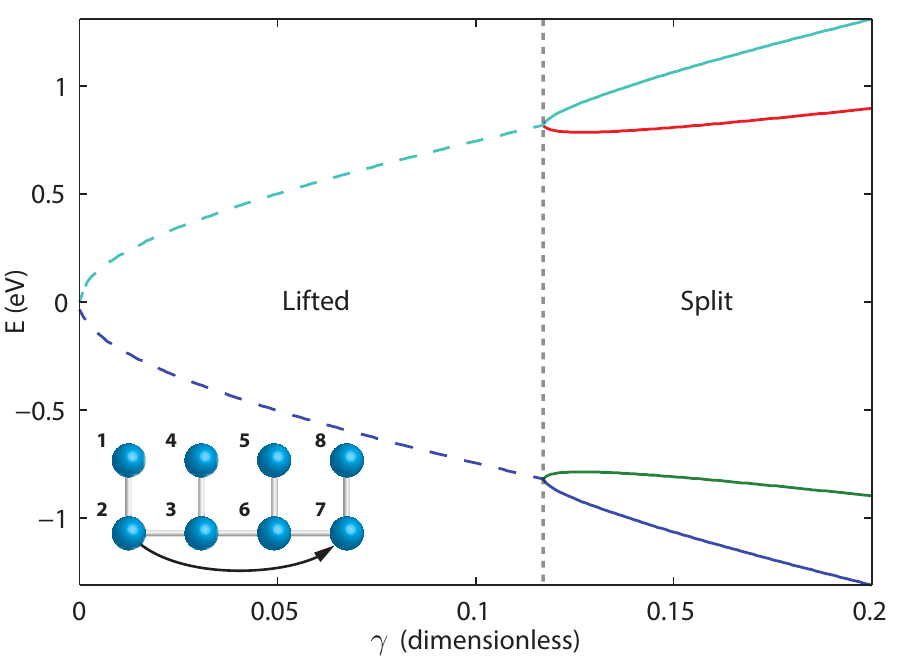}
\caption{The location of the nodes (solid lines) and minima (dashed lines) in the element $G_{27}$ of the retarded Green's function associated with the depicted eight-site extended Hubbard system. Here $\gamma$
is a dimensionless parameter that scales the strength of the interactions. The calculation was performed via exact diagonalization. }
\label{eightSiets_nodeLocationVsInteractionStrength}
\end{figure}

We discuss only briefly cases involving more than four sites. In particular, we consider an eight-site extended Hubbard model depicted
in Figure \ref{eightSiets_nodeLocationVsInteractionStrength} together with its node structure. Here the positions
of local minima (dashed lines) and nodes (solid lines) in the element $G_{27}$ of the retarded Green's function are shown as a function of a
dimensionless prefactor $\gamma$ that premultiplies the interaction matrix $U_{nm}$. When this prefactor is unity, the parameters
in the Hamiltonian are consitent with those for an organic molecule with the appropriate symmetry. 

With $\gamma = 0$ this system exhibits a fourth order degenerate node where $G_{27} \propto E^4$. This splits into two local minima as interactions are introduced. As the strength of interactions passes some critical value $\gamma_c$ these minima bifurcate
and produce two lowest order nodes. Each of these three cases (i.e. $\gamma = 0$, $0 < \gamma < \gamma_c$, and $\gamma > \gamma_c$) can be understood in terms of the properties
of the roots of the polynomial from Eq. \eqref{nodePolynomial}.

\section{The connection between propagation and transmission}
\label{sec:transport}
In section \ref{sec:seriesPropagation} we assumed little about the Green's function except that it vanishes when $t < 0$. In particular, although the
intuitive case of $G_{\beta\alpha}(t, 0) = -i\hbar \Theta(t) \langle d_\beta(t) d^\dagger_\alpha(0) \rangle$ was
considered first to formulate a definition of series propagation, the preceding applies equally to the retarded Green's
function\cite{keldyshPaper} from the Keldysh formalism: $G_{\alpha\beta}(t, 0) = -i\hbar \Theta(t) \langle \{ d_\beta(t) ,
d^\dagger_\alpha(0) \} \rangle$. Transport related quantities are elegantly expressed via this quantity, as it combines
the amplitudes for particle-like and hole-like processes, both of which contribute to e.g. charge and heat transport.

For a nanostructure wherein orbitals $\mu$ and $\nu$ are coupled to a metallic electrode via hopping
matrix elements, the corresponding tunneling self-energy is of the form\cite{jauhoBook, dattaBook} $\Sigma^T_{nm} = \Sigma^T_\mu \delta_{\mu n}\delta_{\mu m}
+ \Sigma^T_\nu \delta_{\nu n}\delta_{\nu m}$. Under these conditions the Dyson equation for the dressed retarded Green's function:
$\mathcal{G}$:\cite{keldyshPaper,dattaBook}
\[
\mathcal{G} = G + G \Sigma^T \mathcal{G}
\]
implies:
\begin{equation}
	 \mathcal{G}_{\mu \nu} = \frac{G_{\mu \nu}}{ (1 - G_{\mu\mu} \Sigma^T_\mu)(1 - G_{\nu\nu}\Sigma^T_\nu) - G_{\mu\nu}\Sigma^T_\nu G_{\nu\mu}\Sigma^T_\mu}.
\end{equation}
Here we have assumed that the full self-energy associated with coupling is equal to the tunneling self-energy,
i.e. taken the so-called elastic cotunneling approximation.\cite{1991groshev, 1990averin,2004glazman,justinManybodyPaper}

In the broad-band limit\cite{meirWingreenTimeDependentReference} $\Sigma^T_\mu = -\frac{i\Gamma^\mu}{2}$ and $\Sigma^T_\nu = -\frac{i\Gamma^\nu}{2}$ in terms of elements $\Gamma_\mu$
and $\Gamma_\nu$ of a tunneling-width matrix. In this case the transmission function between the electrodes coupled to
$\mu$ and $\nu$ is then given by:\cite{landauerLandauerFormula,landauerLandauerFormula2,buttikerFormula,dattaBook}
\begin{equation}
T_{\mu\nu} = \Gamma^\mu \Gamma^\nu |\mathcal{G}_{\mu \nu}|^2 \propto G_{\mu\nu} \nonumber
\end{equation}
With regard to this work, the most important implication
of the foregoing is that a node in $G_{\mu\nu}$ is sufficient for a node in $\mathcal{G}_{\mu\nu}$ and consequently $T_{\mu\nu}$.

\begin{widetext}
\section{A two-pole approximation for the eigenvalues of the retarded Green's function in extended Hubbard models}
\label{sec:spectralDecomposition}

The retarded Green's function of an interacting system can be expressed in terms many-body energies and eigenstates via a Lehmann representation.\cite{lehmannRepresentationPaper1,lehmannRepresentationPaper2,jauhoBook,justinManybodyPaper} In
particular, at zero temperature:
\begin{align}
	\label{exactGnm}
	G_{nm} = \underbrace{ \frac{1}{\Omega_N}\sum_{ {\Psi \in \mathcal{B}^0_N} \atop {\Psi' \in \mathcal{B}_{N + 1}} } \frac{ \langle \Psi | d_n | \Psi' \rangle \langle \Psi' | d^\dagger_m | \Psi \rangle }{E - \left[E_{\Psi'} - E_{\Psi}\right] + i0^+}}_{\textrm{Particle-like propagation}} + \underbrace{\frac{1}{\Omega_N}\sum_{ {\Psi \in \mathcal{B}^0_N} \atop {\Psi' \in \mathcal{B}_{N - 1}} } \frac{\langle \Psi | d^\dagger_m | \Psi' \rangle \langle \Psi' | d_n | \Psi \rangle}{E + \left[E_{\Psi'} - E_{\Psi}\right] + i0^+}}_{\textrm{Hole-like propagation}}
\end{align}
where $\mathcal{B}^0_N = \{|\Psi\rangle\}$ is an orthonormal set of statistically occupied $N$ particle ground states, $\mathcal{B}_{N \pm 1} = \{|\Psi'\rangle\}$ is an energy eigenbasis for
the space of $N \pm 1$ particle states, and $E_{\Psi}$ and $E_{\Psi'}$ are many-body energies. In the present context $d^\dagger_n$ creates a particle on the $n$th spin-orbital
in an extended Hubbard model,
and $\Omega_N$ is the number of statistically accessible $N$ particle ground states. For the sake of brevity, we consider the spin degrees of freedom to be implicit in the index $n$.

As an alternative to working in a localized basis, one can consider creation operators $d^\dagger_\nu(E)$ defined by:
\begin{equation}
	d^\dagger_n = \sum_\nu \langle n | \nu\rangle d^\dagger_\nu(E) 
	\end{equation}
	or equivalently:
	\begin{equation}
		d^\dagger_\nu(E) = \sum_\nu \langle \nu|n\rangle d^\dagger_n
\end{equation}
Here $\{ |\nu(E)\rangle\}$ is an energy-dependent single-particle basis chosen to diagonalize the retarded Green's function, i.e. such that:
\begin{equation}
	G = \sum_\nu \lambda_{\nu}| \nu \rangle \langle \nu |
\end{equation}
where $\lambda_{\nu}(E)$ are the eigenvalues of $G$. This is always possible for an isolated system because the retarded
Green's function is a normal matrix\footnote{I.e. $G$ commutes with its adjoint.} in this case. Without loss
of generality, we also assume that $\{|\nu(E)\rangle\}$ are eigenvectors of a set of generators for all the one-body symmetries of
the system under consideration.

Formally, Eq. \eqref{exactGnm} and the definition of $\{|\nu\rangle\}$ implies:
\begin{align}
	\label{greensFunctionEigenvalue}
\lambda_{\nu} = \underbrace{ \frac{1}{\Omega_N} \sum_{ {\Psi \in \mathcal{B}^0_N} \atop {\Psi' \in \mathcal{B}_{N + 1}} } \frac{ \mathcal{Z}_\nu^{\Psi \rightarrow \Psi'} }{E - \left[E_{\Psi'} - E_{\Psi}\right] + i0^+}}_{\textrm{Particle-like propagation}} + \underbrace{\frac{1}{\Omega_N} \sum_{ {\Psi \in \mathcal{B}^0_N} \atop {\Psi' \in \mathcal{B}_{N - 1}} } \frac{\mathcal{Z}_\nu^{\Psi' \rightarrow \Psi}}{E + \left[E_{\Psi'} - E_{\Psi}\right] + i0^+}}_{\textrm{Hole-like propagation}}
\end{align}
where $\mathcal{Z}_\nu^{\Psi \rightarrow \Psi'} = |\langle \Psi' |d^\dagger_{\nu}|\Psi\rangle|^2$ are spectral weights associated with the transitions
between many-body states. Since all $\Psi \in \mathcal{B}^0_N$ are associated with the same degenerate ground-state energy, this is equivalent to:
\begin{align}
	\label{greensFunctionEigenvalue}
\lambda_{\nu} = \sum_{ \Psi' \in \mathcal{B}_{N + 1} } \frac{ Z_\nu^{\Psi'} }{E - \varepsilon_\nu^{\Psi'} + i0^+} + \sum_{ \Psi' \in \mathcal{B}_{N - 1} } \frac{Z_\nu^{\Psi'}}{E + \varepsilon_\nu^{\Psi'} + i0^+} 
\end{align}
where:
\begin{equation}
Z^{\Psi'}_\nu = \frac{1}{\Omega_N} \sum_{\Psi \in \mathcal{B}^0_N } \mathcal{Z}_\nu^{\Psi \rightarrow \Psi'}
\end{equation}
and:
\begin{equation}
\varepsilon_\nu^{\Psi'} = E_{\Psi'} - E_{\Psi}
\end{equation}

In the sum above, when $\Psi'$ corresponds to a correlation-induced excited state, it contributes a narrow resonance at high-energy, e.g. in
the shaded region inaccessible
at the Fermi level in Figures \ref{twoSitesGreensFunctionPanel} and \ref{fourSitesGreensFunctionPanel}. In the white regions in the same figures,
these can be neglected or, as seen shortly, treated by renormalizing spectral weights. This suggests that, in the absence of orbital degeneracy,
the eigenvalues $\lambda_\nu$ be approximated as:
\begin{align}
	\label{new}
\lambda_{\nu} \approx \frac{ Z^{\text{p}}_\nu }{E - \varepsilon_\nu^{\text{p}} + i0^+} +  \frac{Z^{\text{h}}_\nu}{E + \varepsilon_\nu^{\text{h}} + i0^+} 
\end{align}
where:
\begin{equation}
Z^{\text{p}}_\nu = \frac{1}{\Omega_N}  \sum_{{\Psi \in \mathcal{B}^0_N} \atop {\Psi' \in \mathcal{B}^0_{N + 1}}  } \mathcal{Z}_\nu^{\Psi \rightarrow \Psi'}
\end{equation}
and:
\begin{equation}
Z^{\text{h}}_\nu = \frac{1}{\Omega_N}  \sum_{{\Psi \in \mathcal{B}^0_N} \atop {\Psi' \in \mathcal{B}^0_{N - 1}}  } \mathcal{Z}_\nu^{\Psi \rightarrow \Psi'}
\end{equation}
Here, $\varepsilon_\nu^{\text{p},\text{h}}$ are energies associated with the particle-like and hole-like transitions with nonzero
spectral weight closest to the Fermi level. Due to our choice of $|\nu\rangle$, the foregoing is also valid in the presence of an orbital degeneracy
if it is due to a symmetry not broken by interactions.

To explore the approximation described by Eq. \eqref{new}, it is useful to consider briefly the limit of constant interactions,\cite{konig1997} i.e. $U_{nm} = U$, in which case it is equivalent
to its exact counterpart
Eq. \eqref{greensFunctionEigenvalue}. In this case the many-body
states are Slater determinants and the
eigenvectors $|\nu\rangle$ are just the noninteracting ones. The spectral weights $\mathcal{Z}_\nu^{\Psi \rightarrow \Psi'}$ are then zero or unity depending the occupancy of the
spin-orbital $|\nu\rangle$ in $|\Psi\rangle$ and
$|\Psi'\rangle$. Likewise, the energies $\varepsilon_\nu^{\text{p},\text{h}}$ are given exacly by $\varepsilon^{\text{p}}_\nu = \varepsilon_\nu + U(\left[N - N_0\right] + \frac{1}{2})$ and
$\varepsilon^{\text{h}}_\nu = -\varepsilon_\nu - U(\left[N - N_0\right] - \frac{1}{2})$. There are therefore no correlation induced resonances at high-energies, and the
resonances in Eq. $\eqref{new}$ for $\lambda_\nu$ correspond to the addition or removal of a particle from the single-particle state $|\nu\rangle$.

\begin{figure}
	\centering
	\includegraphics[scale=1.0]{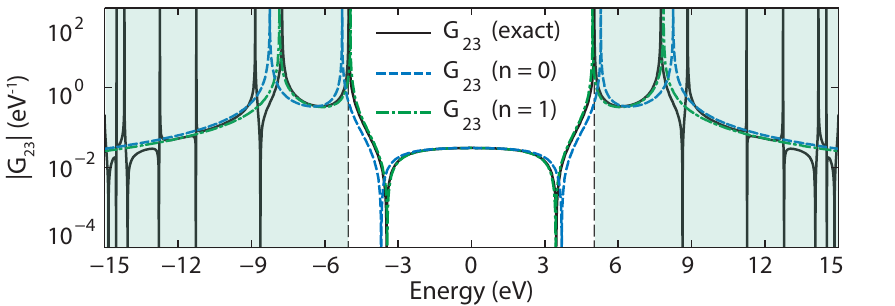}
\caption{The retarded Green's function of the four-site extended Hubbard model considered in section \ref{sec:caseStudies} calculated exactly or using zeroth or first order
approximations for the eigenvalues of the Coulomb self-energy. The eigenvectors are evaluated with the energy fixed at $E = 0$. 
The linear approximation is essentially exact below $E \approx \pm 8$ eV, where the Coulomb self-energy becomes non-analytic due to a singularity associated with a node in an eigenvalue of $G$.
\label{greensFunction_approximation_comparison}}
\end{figure}

If distance-dependent interactions are dialed on, i.e. $U_{nm} = U \longrightarrow U + \Delta U_{nm}$, the eigenvalues $\lambda_{\nu}$ in Eq. \eqref{greensFunctionEigenvalue} and potentially the eigenvectors $\{|\nu\rangle\}$
deviate from those with constant interactions. At low energies (e.g. the white region in Figures \ref{twoSitesGreensFunctionPanel} and \ref{fourSitesGreensFunctionPanel}), this comes primarily from
a shift in the positions of the dominant resonances, which correspond to those present with constant interactions, as well as a reduction of their spectral weights $Z^{\text{p,h}}_\nu$.
The latter occurs as spectral weight is transferred to the correlation-induced resonances at high energy.

Overall, we find that in the simple cases considered herein, an expression of the form \eqref{new} with ``effective'' parameters $\tilde{\varepsilon}^{\text{p},\text{h}}_\nu$, $0 < \tilde{Z}^{\text{p,h}}_\nu < 1$,
and $\tilde{U} > 0$ reproduces the low-energy Green's function with high accuracy. Moreover, with long-ranged interactions these parameters are typically near to their bare values. 
While this method is similar to two-pole approximations applied to the Hubbard model\cite{hubbardPaper,rothPaper}, it is expected
to work best when the length scale of the interactions is larger than or comparable to the size of the system considered.

This approximation may also be cast in terms of the eigenvalues of the Coulomb self-energy $\Sigma_\nu$, which, from the Dyson equation\cite{dattaBook, keldyshPaper}, are related to the eigenvalues above simply by:
\begin{equation} 
	\lambda_\nu = \frac{1}{E - \varepsilon_\nu - \Sigma_\nu}
\end{equation}
where $\varepsilon_\nu$ are the noninteracting energies. 
Eq. \eqref{new} then corresponds to a Laurent expansion of the Coulomb self-energy with the form:
\begin{equation}
	\Sigma_\nu = c_{-1}\left(E - E_{0}\right)^{-1} + c_0 + c_1\left(E - E_{0}\right)
\end{equation}

If there is particle-hole symmetry so that $\tilde{Z}^{\text{p}}_\nu = \tilde{Z}^{\text{h}}_\nu \equiv \frac{1}{2}\tilde{Z}_\nu$, then
$E_0 = \tilde{\varepsilon} \, + \, \tilde{U}\left[N - N_0\right]$, $c_{-1} = \tilde{Z}^{-1}_\nu\left(\frac{\tilde{U}}{2}\right)^2$, $c_0 = E_0 - \varepsilon_\nu $, and $c_{1} = 1 - \tilde{Z}^{-1}_\nu$.
Here $N_0$ is the number of sites in the extended Hubbard Hamiltonian. The residue is associated
 with the nodes in $\lambda_\nu$ and the linear term adjusts the spectral weight of the dominant resonances in $\lambda_\nu$ in a manner reminiscent of a Fermi liquid.\cite{landauFermiLiquidPaper}.

The assumption of particle-hole symmetry above is appropriate when $|\nu\rangle$ is expected to be half-filled. If, for example, $|\nu\rangle$ is instead expected to be completely empty,
then $\tilde{Z}^\textrm{h}_\nu = 0$. In this case $E_0 = \tilde{\varepsilon}_\nu + \tilde{U}(\left[N - N_0\right] + \frac{1}{2})$, $c_{-1} = 0$,
$c_0 = E_0 - \varepsilon_\nu$, and $c_1 = 1 - \tilde{Z}^{-1}_\nu$. In general, this approximation breaks down near correlation-induced nodes in the eigenvalue, which correspond to singularities
in the exact Coulomb self-energy that render it non-analytic.
 
As a concrete example of the foregoing, we offer Figure \ref{greensFunction_approximation_comparison}, which shows the element $G_{23}$ of the Green's function of system $A + B$ (Figure \ref{four_sites}) with
long-ranged Coulomb interactions as specified by Eq. \eqref{initialParameterization}. This is calculated at half-filling using zeroth or
first order approximations for the eigenvalues $\Sigma_\nu$ and with the eigenvectors $|\nu\rangle$ held constant. The linear ($n = 1$) approximation is excellent
below $E \approx \pm 8$ eV, where the Coulomb self-energy becomes non-analytic due to a correlation-induced singularity in one of its eigenvalues. Equivalently,
at this location there is a node in an eigenvalue
of the Green's function.\\

\end{widetext}

\bibliographystyle{unsrt}
\bibliography{series_transmission.bib}

\begin{thebibliography}{10}

\bibitem{justinMottNodePaper}
J.P. Bergfield, G.C. Solomon, C.A. Stafford, and M.A. Ratner.
\newblock Novel quantum interference effects in transport through molecular
  radicals.
\newblock {\em Nano letters}, 11(7):2759--2764, 2011.

\bibitem{hubbardPaper}
J.~Hubbard.
\newblock Electron correlations in narrow energy bands.
\newblock {\em Proceedings of the Royal Society of London. Series A.
  Mathematical and Physical Sciences}, 276(1365):238--257, 1963.

\bibitem{Ohno64}
K.~Ohno.
\newblock Some remarks on the {P}ariser-{P}arr-{P}ople method.
\newblock {\em Theor. Chim. Acta}, 2:219, 1964.

\bibitem{Castleton02}
C.~W.~M. {Castleton} and W.~{Barford}.
\newblock Screening and the quantitative pi-model description of the optical
  spectra and polarizations of phenyl based oligomers.
\newblock {\em J. Chem. Phys.}, 117:3570--3582, 2002.

\bibitem{barrPiEFTPaper}
J.~D. Barr, C.~A. Stafford, and J.~P. Bergfield.
\newblock Effective field theory of interacting pi-electrons.
\newblock {\em Phys. Rev. B}, 86:115403, Sep 2012.

\bibitem{dattaMasterEquation}
B.~Muralidharan, A.~W. Ghosh, and S.~Datta.
\newblock Probing electronic excitations in molecular conduction.
\newblock {\em Phys. Rev. B}, 73:155410, Apr 2006.

\bibitem{rinconManybodyTransportAnnulenes}
Juli\'an Rinc\'on, K.~Hallberg, A.~A. Aligia, and S.~Ramasesha.
\newblock Quantum interference in coherent molecular conductance.
\newblock {\em Phys. Rev. Lett.}, 103:266807, Dec 2009.

\bibitem{justinManybodyPaper}
J.~P. Bergfield and C.~A. Stafford.
\newblock Many-body theory of electronic transport in single-molecule
  heterojunctions.
\newblock {\em Phys. Rev. B}, 79:245125, Jun 2009.

\bibitem{justinSupernodePaper}
Justin~P. Bergfield, Michelle~A. Solis, and Charles~A. Stafford.
\newblock Giant thermoelectric effect from transmission supernodes.
\newblock {\em ACS Nano}, 4(9):5314--5320, 2010.

\bibitem{2007bohr}
Dan Bohr, Peter Schmitteckert, and Peter W{\"o}lfle.
\newblock Dmrg evaluation of the kubo formula—conductance of strongly
  interacting quantum systems.
\newblock {\em EPL (Europhysics Letters)}, 73(2):246, 2007.

\bibitem{meirWingreenPaper}
Yigal Meir and Ned~S. Wingreen.
\newblock Landauer formula for the current through an interacting electron
  region.
\newblock {\em Phys. Rev. Lett.}, 68:2512--2515, Apr 1992.

\bibitem{meirWingreenPaper2}
Yigal Meir, Ned~S. Wingreen, and Patrick~A. Lee.
\newblock Transport through a strongly interacting electron system: Theory of
  periodic conductance oscillations.
\newblock {\em Phys. Rev. Lett.}, 66:3048--3051, Jun 1991.

\bibitem{meirWingreenTimeDependentReference}
Antti-Pekka Jauho, Ned~S. Wingreen, and Yigal Meir.
\newblock Time-dependent transport in interacting and noninteracting
  resonant-tunneling systems.
\newblock {\em Phys. Rev. B}, 50:5528--5544, Aug 1994.

\bibitem{BDTNodePaper1}
P.~Sautet and C.~Joachim.
\newblock Electronic interference produced by a benzene embedded in a
  polyacetylene chain.
\newblock {\em Chemical Physics Letters}, 153(6):511 -- 516, 1988.

\bibitem{BDTNodePaper2}
Cendrine Patoux, Christophe Coudret, Jean-Pierre Launay, Christian Joachim, and
  André Gourdon.
\newblock Topological effects on intramolecular electron transfer via quantum
  interference.
\newblock {\em Inorganic Chemistry}, 36(22):5037--5049, 1997.

\bibitem{BDTNodePaper3}
S.~N. YALIRAKI and MARK~A. RATNER.
\newblock Interplay of topology and chemical stability on the electronic
  transport of molecular junctions.
\newblock {\em Annals of the New York Academy of Sciences}, 960(1):153--162,
  2002.

\bibitem{BDTNodePaper4}
R~Stadler, S~Ami, C~Joachim, and M~Forshaw.
\newblock Integrating logic functions inside a single molecule.
\newblock {\em Nanotechnology}, 15(4):S115, 2004.

\bibitem{BDTNodePaper5}
Derek Walter, Daniel Neuhauser, and Roi Baer.
\newblock Quantum interference in polycyclic hydrocarbon molecular wires.
\newblock {\em Chemical Physics}, 299(1):139 -- 145, 2004.

\bibitem{BDTNodePaper8}
San-Huang Ke, Weitao Yang, and Harold~U. Baranger.
\newblock Quantum-interference-controlled molecular electronics.
\newblock {\em Nano Letters}, 8(10):3257--3261, 2008.
\newblock PMID: 18803424.

\bibitem{BDTNodePaper9}
Gemma~C. Solomon, David~Q. Andrews, Richard~P. Van~Duyne, and Mark~A. Ratner.
\newblock Electron transport through conjugated molecules: When the π system
  only tells part of the story.
\newblock {\em ChemPhysChem}, 10(1):257--264, 2009.

\bibitem{BDTNodePaper10}
Thorsten Hansen, Gemma~C. Solomon, David~Q. Andrews, and Mark~A. Ratner.
\newblock Interfering pathways in benzene: An analytical treatment.
\newblock {\em The Journal of Chemical Physics}, 131(19):194704, 2009.

\bibitem{BDTNodePaper11}
Aleksey~A. Kocherzhenko, Ferdinand~C. Grozema, and Laurens D.~A. Siebbeles.
\newblock Charge transfer through molecules with multiple pathways: Quantum
  interference and dephasing.
\newblock {\em The Journal of Physical Chemistry C}, 114(17):7973--7979, 2010.

\bibitem{BDTNodePaper12}
Marcel Mayor, Heiko~B. Weber, Joachim Reichert, Mark Elbing, Carsten von
  Hänisch, Detlef Beckmann, and Matthias Fischer.
\newblock Electric current through a molecular rod—relevance of the position
  of the anchor groups.
\newblock {\em Angewandte Chemie International Edition}, 42(47):5834--5838,
  2003.

\bibitem{BDTNodePaper13}
Manabu Kiguchi, Hisao Nakamura, Yuuta Takahashi, Takuya Takahashi, and
  Tatsuhiko Ohto.
\newblock Effect of anchoring group position on formation and conductance of a
  single disubstituted benzene molecule bridging au electrodes: Change of
  conductive molecular orbital and electron pathway.
\newblock {\em The Journal of Physical Chemistry C}, 114(50):22254--22261,
  2010.

\bibitem{connectivityConductanceExperimentPaper}
M.~Mayor, H.B. Weber, J.~Reichert, M.~Elbing, C.~von H{\"a}nisch, D.~Beckmann,
  and M.~Fischer.
\newblock Electric current through a molecular rod—relevance of the position
  of the anchor groups.
\newblock {\em Angewandte Chemie International Edition}, 42(47):5834--5838,
  2003.

\bibitem{justinNodeSplittingPaper}
G.C. Solomon, J.P. Bergfield, C.A. Stafford, and M.A. Ratner.
\newblock When “small” terms matter: Coupled interference features in the
  transport properties of cross-conjugated molecules.
\newblock {\em Beilstein Journal of Nanotechnology}, 2:862, 2011.

\bibitem{Note1}
After preparation of this manuscript, we became aware via private
  correspondence that J. Bergfield has recently observed node splitting within
  a Hubbard model of a molecular junction in unpublished work.

\bibitem{rothPaper}
Laura~M. Roth.
\newblock Electron correlation in narrow energy bands. i. the two-pole
  approximation in a narrow $s$ band.
\newblock {\em Phys. Rev.}, 184:451--459, Aug 1969.

\bibitem{landauFermiLiquidPaper}
LD~Landau.
\newblock The theory of a fermi liquid.
\newblock {\em Sov. Phys. JETP}, 3(6):920, 1957.

\bibitem{andersonAndersonModel}
P.~W. Anderson.
\newblock Localized magnetic states in metals.
\newblock {\em Phys. Rev.}, 124:41--53, Oct 1961.

\bibitem{mottNodePaper}
T.D. Stanescu, P.~Phillips, and T.P. Choy.
\newblock Theory of the luttinger surface in doped mott insulators.
\newblock {\em Physical Review B}, 75(10):104503, 2007.

\bibitem{1948feynman}
R.~P. Feynman.
\newblock Space-time approach to non-relativistic quantum mechanics.
\newblock {\em Rev. Mod. Phys.}, 20:367--387, Apr 1948.

\bibitem{dysonEquationPaper}
F.J. Dyson.
\newblock The s matrix in quantum electrodynamics.
\newblock {\em Physical Review}, 75(11):1736, 1949.

\bibitem{keldyshPaper}
L.V. Keldysh.
\newblock Diagram technique for nonequilibrium processes.
\newblock {\em Zh. Eksp. Teor. Fiz}, 47(4):151--165, 1964.

\bibitem{jauhoBook}
H.~Haug and A.P. Jauho.
\newblock {\em Quantum Kinetics in Transport and Optics of Semiconductors}.
\newblock Springer Series in Solid-state Sciences, v. 123. Springer-Verlag
  Berlin Heidelberg, 2008.

\bibitem{1991groshev}
A.~Groshev, T.~Ivanov, and V.~Valtchinov.
\newblock Charging effects of a single quantum level in a box.
\newblock {\em Phys. Rev. Lett.}, 66:1082--1085, Feb 1991.

\bibitem{1990averin}
DV~Averin and Yu~V Nazarov.
\newblock Virtual electron diffusion during quantum tunneling of the electric
  charge.
\newblock {\em Physical review letters}, 65(19):2446--2449, 1990.

\bibitem{2004glazman}
Michael Pustilnik and Leonid Glazman.
\newblock Kondo effect in quantum dots.
\newblock {\em Journal of Physics: Condensed Matter}, 16(16):R513, 2004.

\bibitem{wilkinsonPaper}
JH~Wilkinson.
\newblock The evaluation of the zeros of ill-conditioned polynomials. part i.
\newblock {\em Numerische Mathematik}, 1(1):150--166, 1959.

\bibitem{ohnoPaper}
Kimio Ohno.
\newblock Some remarks on the pariser-parr-pople method.
\newblock {\em Theoretical Chemistry Accounts: Theory, Computation, and
  Modeling (Theoretica Chimica Acta)}, 2:219--227, 1964.
\newblock 10.1007/BF00528281.

\bibitem{castletonPaper}
C.~W.~M. Castleton and W.~Barford.
\newblock Screening and the quantitative pi-model description of the optical
  spectra and polarizations of phenyl based oligomers.
\newblock {\em The Journal of Chemical Physics}, 117(8):3570--3582, 2002.

\bibitem{Note2}
Room temperature calculations were carried out and yielded results visually
  indistinguishable from those presented here.

\bibitem{dattaBook}
S.~Datta.
\newblock {\em Electronic Transport in Mesoscopic Systems}.
\newblock Cambridge Studies in Semiconductor Physics and Microelectronic
  Engineering. Cambridge University Press, 1997.

\bibitem{luttingerTheorem1}
JM~Luttinger.
\newblock Fermi surface and some simple equilibrium properties of a system of
  interacting fermions.
\newblock {\em Physical Review}, 119(4):1153, 1960.

\bibitem{luttingerTheorem2}
J.~M. Luttinger and J.~C. Ward.
\newblock Ground-state energy of a many-fermion system. ii.
\newblock {\em Phys. Rev.}, 118:1417--1427, Jun 1960.

\bibitem{averinCoulombBlockadeTheory}
DV~Averin and KK~Likharev.
\newblock Coulomb blockade of single-electron tunneling, and coherent
  oscillations in small tunnel junctions.
\newblock {\em Journal of low temperature physics}, 62(3):345--373, 1986.

\bibitem{fultonCoulombBlockade}
T.~A. Fulton and G.~J. Dolan.
\newblock Observation of single-electron charging effects in small tunnel
  junctions.
\newblock {\em Phys. Rev. Lett.}, 59:109--112, Jul 1987.

\bibitem{giaeverNascentCoulombBlockade}
I.~Giaever and H.~R. Zeller.
\newblock Superconductivity of small tin particles measured by tunneling.
\newblock {\em Phys. Rev. Lett.}, 20:1504--1507, Jun 1968.

\bibitem{giaeverNascentCoulombBlockade2}
H.~R. Zeller and I.~Giaever.
\newblock Tunneling, zero-bias anomalies, and small superconductors.
\newblock {\em Phys. Rev.}, 181:789--799, May 1969.

\bibitem{Note3}
More precisely, associated with the same eigenvalue of $G$; the notion of a
  single-particle orbital here is formally justifiable when $U_{nm}$ is
  constant.

\bibitem{Note4}
To be precise, nonperturbative effects are included as well; therefore, in this
  context the diagrams mentioned should be viewed as a conceptual device used
  to classify processes. The only calculation performed diagrammatically was
  self-consistent Hartree-Fock.

\bibitem{landauerLandauerFormula}
Rolf Landauer.
\newblock Spatial variation of currents and fields due to localized scatterers
  in metallic conduction.
\newblock {\em IBM Journal of Research and Development}, 1(3):223--231, 1957.

\bibitem{landauerLandauerFormula2}
Rolf Landauer.
\newblock Electrical resistance of disordered one-dimensional lattices.
\newblock {\em Philosophical Magazine}, 21(172):863--867, 1970.

\bibitem{buttikerFormula}
Markus B{\"u}ttiker.
\newblock Absence of backscattering in the quantum hall effect in multiprobe
  conductors.
\newblock {\em Physical Review B}, 38(14):9375, 1988.

\bibitem{lehmannRepresentationPaper1}
G.~Kallen.
\newblock On the definition of the renormalization constants in quantum
  electrodynamics.
\newblock {\em Helvetica Physica Acta (Switzerland)}, 25, 1952.

\bibitem{lehmannRepresentationPaper2}
H.~Lehmann, K.~Symanzik, and W.~Zimmermann.
\newblock On the formulation of quantized field theories—ii.
\newblock {\em Il Nuovo Cimento (1955-1965)}, 6(2):319--333, 1957.

\bibitem{Note5}
I.e. $G$ commutes with its adjoint.

\bibitem{konig1997}
J\"urgen K\"onig, Herbert Schoeller, and Gerd Sch\"on.
\newblock Cotunneling at resonance for the single-electron transistor.
\newblock {\em Phys. Rev. Lett.}, 78:4482--4485, Jun 1997.

\end{thebibliography}

\end{document}